\documentclass{amsproc}
\usepackage{graphicx,mathrsfs}
\usepackage{tikz}
\usepackage{stmaryrd}
\usepackage{booktabs}
\usepackage{tipa}
\theoremstyle{definition}

\theoremstyle{remark}

\numberwithin{equation}{section}

\def\m@th{\mathsurround=0pt}
\mathchardef\bracell="0365
\def\upbrall{$\m@th\bracell$}
\def\undertilde#1{\mathop{\vtop{\ialign{##\crcr
   $\hfil\displaystyle{#1}\hfil$\crcr
    \noalign
    {\kern1.5pt\nointerlineskip}
    \upbrall\crcr\noalign{\kern1pt
  }}}}\limits}

\mathchardef\hatbracell="0362
\def\hatupbrall{$\m@th\hatbracell$}
\def\underhat#1{\mathop{\vtop{\ialign{##\crcr
   $\hfil\displaystyle{#1}\hfil$\crcr
    \noalign
    {\kern1.5pt\nointerlineskip} 
    \hatupbrall\crcr\noalign{\kern1pt 
  }}}}\limits}

\begin{document}

\title{Discrete Painlev\'e equations and their Lax pairs as reductions of integrable lattice equations}

\author{C. M. Ormerod}

\address{C. M. Ormerod \\ La Trobe University \\ Bundoora, VIC, 3086 \\ Australia}
\email{C.Ormerod@latrobe.edu.au}

\author{Peter H. van der Kamp}

\address{Peter H. van der Kamp \\ La Trobe University \\ Bundoora, VIC, 3086 \\ Australia}
\email{P.vanderKamp@latrobe.edu.au}

\author{G.R.W. Quispel}

\address{G.R.W. Quispel\\ La Trobe University \\ Bundoora, VIC, 3086 \\ Australia}
\email{R.Quispel@latrobe.edu.au}

\begin{abstract}
We describe a method to obtain Lax pairs for periodic reductions of a rather general class of integrable non-autonomous lattice equations. The method is applied to obtain reductions of the non-autonomous discrete Korteweg-de Vries equation and non-autonomous discrete Schwarzian Korteweg-de Vries equation, which yield a discrete analogue of the fourth Painlev\'e equation, a $q$-analogue of the sixth Painlev\'e equation and the $q$-Painlev\'e equation with a symmetry group of affine Weyl type $E_6^{(1)}$.
\end{abstract}


\maketitle

Integrable partial difference equations are discrete time and discrete space analogues of integrable partial differential equations, which often admit classical integrable partial differential equations as continuum limits \cite{AblowitzLadik:Sols, Hirota:DKdV, Nijhoff:lkdvreview, QNCvdL:NSG}. Integrable ordinary difference equations are discrete analogues of integrable ordinary differential equations. Integrable ordinary difference equations admit integrable ordinary differential equations as continuum limits \cite{PainleveProperty}. Integrable ordinary and partial difference equations possess discrete analogues of many of the properties associated to the integrability of their continuous counterparts \cite{PainleveProperty, Gramani:Isomonodromic, Omar, Dinh}.

We consider partial difference equations whose evolution on a lattice of points, $w_{l,m}$, is determined by the equation \begin{equation}\label{Q}
Q(w_{l,m},w_{l+1,m},w_{l,m+1},w_{l+1,m+1};\alpha,\beta) = 0,
\end{equation}
where $\alpha$ and $\beta$ are parameters associated with the horizontal and vertical edges respectively. The equation is imposed on each square on the space of independent variables, $(l,m) \in \mathbb{Z}^2$. From a suitable staircase of initial conditions \cite{Quispel:Intreds}, one may determine $w_{l,m}$ for all $(l,m) \in \mathbb{Z}^2$. Imposing the periodic constraint, that 
\begin{equation}\label{periodicity}
w_{l+s_1,m+s_2} = w_{l,m},
\end{equation}
defines a periodic reduction \cite{Quispel:Intreds, SimilarityReds, VanderKamp:IVPs}. We will assume for simplicity that $s_1$ and $s_2$ are both positive. In an analogous way to how similarity reductions of partial differential equations yield ordinary {\em differential} equations \cite{Kruskal}, periodic reductions given by \eqref{periodicity} yield ordinary {\em difference} equations \cite{Gramani:Reductions, dKdVreds, Quispel:Intreds, VanderKamp:IVPs}.

Given a partial differential equation with some similarity reduction, there is a procedure that allows one to obtain a Lax representation of the resulting ordinary differential equation from the Lax representation of the partial differential equation. This holds for autonomous and non-autonomous reductions \cite{FlaschkaNewell}. The discrete analogue of this procedure is fairly straightforward for autonomous reductions \cite{Omar, Dinh, Duality}, however, there has been no direct method for determining the Lax representation for non-autonomous reductions \cite{Hay:Hierarchies, Hay}.

Given a reduction, another task is to determine whether the reduction is a known system of difference equations. For autonomous reductions, one may be able to find a certain parameterisation which identifies the system as a known QRT mapping \cite{QRT1, QRT2}, which may be classified in terms of elliptic surfaces \cite{Duistermaat}. For nonautonomous reductions, one may be able find a parameterisation of the equation that identifies the system as one of the Painlev\'e equations, which are classified by the group of symmetries of their surface of initial conditions \cite{Sakai:Rational}.

The aim of this note is to demonstrate a method, which we outline in \S 1, by which we may directly obtain a Lax representation of both autonomous and  non-autonomous reductions from a Lax representation of partial difference equations in an algorithmic manner. The method gives Lax representations in a manner that is general and concise enough to directly provide the Lax integrability of entire hierarchies of reductions. As an application of this method, we present a reduction of the non-autonomous discrete Schwarzian Korteweg-de Vries equation (which is a non-autonomous version of $Q_1^{\delta=0}$ in the classification of Adler, Bobenko and Suris \cite{ABS:ListI, ABS:ListII}) to the $q$-Painlev\'e equation with $E_6^{(1)}$ symmetry, which is associated with a surface with $A_2^{(1)}$ symmetry (or $q$-$\mathrm{P}(A_2^{(1)})$):
\begin{subequations}\label{qPE6}
\begin{align}
\label{ybar}\left(y' z-1\right) \left(y' z'-1\right)=&\frac{\left(a_1 y'-1\right) \left(a_2 y'-1\right) \left(a_3 y'-1\right) \left(a_4 y'-1\right)}{\left(b_1 q^4 t y'-1\right) \left(b_2 q^4 t y'-1\right)},\\
\label{zbar}(y z-1) \left(y' z-1\right)=&\frac{\theta _1 \left(z-a_1\right) \left(z-a_2\right) \left(z-a_3\right) \left(z-a_4\right)}{\left(b_1 b_2 t z+\theta _1\right) \left(a_1 a_2 a_3 a_4+\theta _1 q^4 t z\right)},
\end{align}
\end{subequations}
where $t' = q^4t$, the $a_i$, $b_i$ and $\theta_1$ are fixed parameters and $q$ is some complex number whose modulus is not $1$. This is the $q$-Painlev\'e equation whose group of B\"acklund transformations is an affine Weyl group of type $E_6^{(1)}$ \cite{Sakai:Rational}.

Our method stands in contrast to two methods of performing reductions of partial difference equations in the literature, namely the method of Hydon et al. \cite{RasinHydon:SymmetryReductions} which is based on the existence of certain Lie point symmetries, and the method of Grammaticos and Ramani, who perform autonomous reductions, then deautonomize the equation via singularity confinement \cite{Gramani:Q4reduction}. While the first method seems to rely on a similar approach to ours, neither method gives rise to the associated linear problem for the reduced equation. The approach most similar to our method has been discussed by Hay et al. \cite{Hay}, in which the form of the monodromy matrix for autonomous reductions, and its properties, are used as an ansatz for an associated linear problem of the non-autonomous reductions of the lattice modified Korteweg-de Vries equation. A further extension to this work successfully determined the associated linear problem for a hierarchy of systems \cite{Hay:Hierarchies}. 

To demonstrate our method we first provide some simple examples in \S 2. We present an autonomous reduction of the discrete potential Korteweg-de Vries equation (dKdV) \cite{Nijhoff:lkdvreview}, then present the non-autonomous generalization of this example. In \S 3 we first present the $q$-Painlev\'e equation associated with the $A_3^{(1)}$ surface (otherwise known as $q$-$\mathrm{P}_{VI}$ \cite{Sakai:qP6}) as a reduction of \eqref{dSlKdV} before going to the higher case where we present the above-mentioned reduction of \eqref{dSlKdV} to \eqref{qPE6}, which we believe to be the first known reduction to this equation.

\section{The method}

We start by imposing \eqref{periodicity} as a constraint on our initial conditions, then the periodicity gives us that there are $s_1 + s_2$ independent initial conditions to define. We solve this periodicity constraint by a specific labelling following \cite{Quispel:Intreds, VanderKamp:IVPs}; let $s_1 = a g$ and $s_2 = bg$ where $g = \mathrm{gcd}(s_1,s_2)$, then the direction of the generating shift, $(c,d)$, associated with the increment $n \to n+1$ is chosen so that
\[
\det \begin{pmatrix} a & b \\ c & d \end{pmatrix} = 1.
\]
We specify an $n \in \mathbb{Z}$ and a $p \in \mathbb{Z}_g$ by letting
\[
n = \det \begin{pmatrix} a & b \\ 
l & m \end{pmatrix}, \hspace{1cm} p \equiv \det \begin{pmatrix} l & m \\ c & d \end{pmatrix} \mod g,
\]
where the labelling of variables is specified by
\begin{equation}\label{labelling}
w_{l,m} \mapsto w_n^{p}.
\end{equation}
In the case in which $g = 1$ the superscript will be omitted. The reduction in the autonomous case is a system of $g$ equations given by
\begin{equation}\label{autred}
Q(w_n^p, w_{n-b}^{p+d}, w_{n+a}^{p-c}, w_{n+a-b}^{p-c+d};\alpha, \beta) = 0, \hspace{1cm} p = 0,1, \ldots, g-1,
\end{equation}
where $\alpha$ and $\beta$ are constants. In the nonautonomous setting, we have
\begin{equation}\label{nautred}
Q(w_n^p, w_{n-b}^{p+d}, w_{n+a}^{p-c}, w_{n+a-b}^{p-c+d};\alpha_l, \beta_m) = 0, \hspace{1cm} p = 0,1, \ldots, g-1,
\end{equation}
where $\alpha_l$ and $\beta_m$ will be, a posteriori, constrained functions of $l$ and $m$. We will now outline how to obtain Lax representations for the autonomous and nonautonomous reductions respectively.

\subsection{Autonomous reductions}

It is known that multilinear partial difference equations that are consistent around a cube are, in a sense, their own Lax pair \cite{NijhoffQ4Lax, BobSur:IntQuad, Birdgman}. For a generic multilinear equation, \eqref{Q}, that is consistent around a cube, a Lax pair may be written as 
\begin{subequations}\label{LaxForm}
\begin{align}
\phi_{l+1,m} = L_{l,m} \phi_{l,m},\\
\phi_{l,m+1} = M_{l,m} \phi_{l,m},
\end{align}
\end{subequations}
where
\begin{subequations}
\begin{align}
L_{l,m} &= \left.\lambda_{l,m} \begin{pmatrix} -\dfrac{\partial Q(x,u,v,0;\alpha,\gamma )}{\partial v} & -Q(x,u,0,0;\alpha,\gamma ) \\
 \dfrac{\partial^2 Q(x,u,v,y;\alpha,\gamma )}{\partial v \partial y} & \dfrac{\partial Q(x,u,0,y;\alpha ,\gamma )}{\partial y}\end{pmatrix}\right|_{\begin{array}{c} x= w_{l,m}\\ u = w_{l+1,m}\end{array}},\\
M_{l,m} &= \left. \mu_{l,m} \begin{pmatrix} -\dfrac{\partial Q(x,u,v,0;\beta,\gamma )}{\partial u} & -Q(x,0,v,0;\beta,\gamma ) \\
 \dfrac{\partial^2 Q(x,u,v,y;\beta,\gamma )}{\partial u \partial y} & \dfrac{\partial Q(x,0,v,y;\beta ,\gamma )}{\partial y}\end{pmatrix} \right|_{\begin{array}{c} x= w_{l,m}\\ v = w_{l,m+1}\end{array}},
\end{align}
\end{subequations}
where $\gamma$ is a spectral parameter. The compatibility condition is
\begin{equation}\label{comp}
M_{l+1,m}L_{l,m} = L_{l,m+1}M_{l,m},
\end{equation}
forcing the prefactors, $\lambda_{l,m}$ and $\mu_{l,m}$, to be chosen in a manner that satisfies the equation
\[
\dfrac{\det L_{l,m+1}}{\det L_{l,m}} = \dfrac{\det M_{l+1,m}}{\det M_{l,m}}.
\]
When the prefactors are appropriately chosen, imposing \eqref{comp} is equivalent to \eqref{Q}. In practice, it is often computationally convenient to deal with some transformation of this Lax pair.

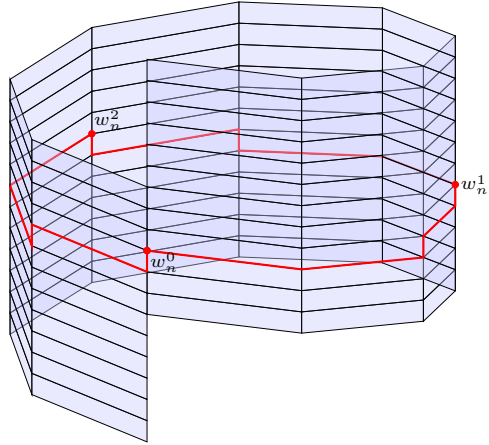
\begin{figure}[!ht]
\begin{tikzpicture}[scale=.6] 
\tikzstyle{conefill} = [fill=blue!20,fill opacity=0.4,very thin]

\filldraw[conefill] (3.34659,0.699536)--(0.166162,0.830568)--(0.166162,1.30181)--(3.34659,1.17077)-- cycle;
\filldraw[conefill] (3.34659,1.17077)--(0.166162,1.30181)--(0.166162,1.77305)--(3.34659,1.64201)-- cycle;
\filldraw[conefill] (3.34659,1.64201)--(0.166162,1.77305)--(0.166162,2.24428)--(3.34659,2.11325)-- cycle;
\filldraw[conefill] (3.34659,2.11325)--(0.166162,2.24428)--(0.166162,2.71552)--(3.34659,2.58449)-- cycle;
\filldraw[conefill] (3.34659,2.58449)--(0.166162,2.71552)--(0.166162,3.18676)--(3.34659,3.05573)-- cycle;
\filldraw[conefill] (3.34659,3.05573)--(0.166162,3.18676)--(0.166162,3.658)--(3.34659,3.52697)-- cycle;
\filldraw[conefill] (3.34659,3.52697)--(0.166162,3.658)--(0.166162,4.12924)--(3.34659,3.99821)-- cycle;
\filldraw[conefill] (3.34659,3.99821)--(0.166162,4.12924)--(0.166162,4.60048)--(3.34659,4.46945)-- cycle;
\filldraw[conefill] (3.34659,4.46945)--(0.166162,4.60048)--(0.166162,5.07172)--(3.34659,4.94069)-- cycle;
\filldraw[conefill] (3.34659,4.94069)--(0.166162,5.07172)--(0.166162,5.54296)--(3.34659,5.41193)-- cycle;
\filldraw[conefill] (3.34659,5.41193)--(0.166162,5.54296)--(0.166162,6.0142)--(3.34659,5.88316)-- cycle;
\filldraw[conefill] (3.34659,5.88316)--(0.166162,6.0142)--(0.166162,6.48543)--(3.34659,6.3544)-- cycle;
\filldraw[conefill] (0.166162,0.830568)--(-3.09201,0.26322)--(-3.09201,0.734459)--(0.166162,1.30181)-- cycle;
\filldraw[conefill] (0.166162,1.30181)--(-3.09201,0.734459)--(-3.09201,1.2057)--(0.166162,1.77305)-- cycle;
\filldraw[conefill] (0.166162,1.77305)--(-3.09201,1.2057)--(-3.09201,1.67694)--(0.166162,2.24428)-- cycle;
\filldraw[conefill] (0.166162,2.24428)--(-3.09201,1.67694)--(-3.09201,2.14818)--(0.166162,2.71552)-- cycle;
\filldraw[conefill] (0.166162,2.71552)--(-3.09201,2.14818)--(-3.09201,2.61941)--(0.166162,3.18676)-- cycle;
\filldraw[conefill] (0.166162,3.18676)--(-3.09201,2.61941)--(-3.09201,3.09065)--(0.166162,3.658)-- cycle;
\filldraw[conefill] (0.166162,3.658)--(-3.09201,3.09065)--(-3.09201,3.56189)--(0.166162,4.12924)-- cycle;
\filldraw[conefill] (0.166162,4.12924)--(-3.09201,3.56189)--(-3.09201,4.03313)--(0.166162,4.60048)-- cycle;
\filldraw[conefill] (0.166162,4.60048)--(-3.09201,4.03313)--(-3.09201,4.50437)--(0.166162,5.07172)-- cycle;
\filldraw[conefill] (0.166162,5.07172)--(-3.09201,4.50437)--(-3.09201,4.97561)--(0.166162,5.54296)-- cycle;
\filldraw[conefill] (0.166162,5.54296)--(-3.09201,4.97561)--(-3.09201,5.44685)--(0.166162,6.0142)-- cycle;
\filldraw[conefill] (0.166162,6.0142)--(-3.09201,5.44685)--(-3.09201,5.91809)--(0.166162,6.48543)-- cycle;
\filldraw[conefill] (-3.09201,0.26322)--(-4.9034,-0.884036)--(-4.9034,-0.412797)--(-3.09201,0.734459)-- cycle;
\filldraw[conefill] (-3.09201,0.734459)--(-4.9034,-0.412797)--(-4.9034,0.0584415)--(-3.09201,1.2057)-- cycle;
\filldraw[conefill] (-3.09201,1.2057)--(-4.9034,0.0584415)--(-4.9034,0.52968)--(-3.09201,1.67694)-- cycle;
\filldraw[conefill] (-3.09201,1.67694)--(-4.9034,0.52968)--(-4.9034,1.00092)--(-3.09201,2.14818)-- cycle;
\filldraw[conefill] (-3.09201,2.14818)--(-4.9034,1.00092)--(-4.9034,1.47216)--(-3.09201,2.61941)-- cycle;
\filldraw[conefill] (-3.09201,2.61941)--(-4.9034,1.47216)--(-4.9034,1.9434)--(-3.09201,3.09065)-- cycle;
\filldraw[conefill] (-3.09201,3.09065)--(-4.9034,1.9434)--(-4.9034,2.41464)--(-3.09201,3.56189)-- cycle;
\filldraw[conefill] (-3.09201,3.56189)--(-4.9034,2.41464)--(-4.9034,2.88587)--(-3.09201,4.03313)-- cycle;
\filldraw[conefill] (-3.09201,4.03313)--(-4.9034,2.88587)--(-4.9034,3.35711)--(-3.09201,4.50437)-- cycle;
\filldraw[conefill] (-3.09201,4.50437)--(-4.9034,3.35711)--(-4.9034,3.82835)--(-3.09201,4.97561)-- cycle;
\filldraw[conefill] (-3.09201,4.97561)--(-4.9034,3.82835)--(-4.9034,4.29959)--(-3.09201,5.44685)-- cycle;
\filldraw[conefill] (-3.09201,5.44685)--(-4.9034,4.29959)--(-4.9034,4.77083)--(-3.09201,5.91809)-- cycle;

\begin{scope}[yshift=.475cm]
\draw[thick,red] (4.96111,1.96339)-- (3.34659,2.58449)--(0.166162,2.71552) -- (0.166162,3.18676)--(-3.09201,2.61941)--(-3.09201,3.09065) -- (-4.9034,1.9434) ;
\end{scope}
\filldraw[conefill] (-4.9034,-0.884036)--(-4.42043,-2.22139)--(-4.42043,-1.75015)--(-4.9034,-0.412797)-- cycle;

\filldraw[conefill] (-4.9034,-0.412797)--(-4.42043,-1.75015)--(-4.42043,-1.27891)--(-4.9034,0.0584415)-- cycle;
\filldraw[conefill] (-4.9034,0.0584415)--(-4.42043,-1.27891)--(-4.42043,-0.80767)--(-4.9034,0.52968)-- cycle;
\filldraw[conefill] (-4.9034,0.52968)--(-4.42043,-0.80767)--(-4.42043,-0.336431)--(-4.9034,1.00092)-- cycle;
\filldraw[conefill] (-4.9034,1.00092)--(-4.42043,-0.336431)--(-4.42043,0.134808)--(-4.9034,1.47216)-- cycle;
\filldraw[conefill] (-4.9034,1.47216)--(-4.42043,0.134808)--(-4.42043,0.606046)--(-4.9034,1.9434)-- cycle;
\filldraw[conefill] (-4.9034,1.9434)--(-4.42043,0.606046)--(-4.42043,1.07729)--(-4.9034,2.41464)-- cycle;
\filldraw[conefill] (-4.9034,2.41464)--(-4.42043,1.07729)--(-4.42043,1.54852)--(-4.9034,2.88587)-- cycle;
\filldraw[conefill] (-4.9034,2.88587)--(-4.42043,1.54852)--(-4.42043,2.01976)--(-4.9034,3.35711)-- cycle;
\filldraw[conefill] (-1.8691,-0.442635)--(1.55681,-0.858957)--(1.55681,-0.387718)--(-1.8691,0.0286035)-- cycle;
\filldraw[conefill] (-1.8691,0.0286035)--(1.55681,-0.387718)--(1.55681,0.0835205)--(-1.8691,0.499842)-- cycle;
\filldraw[conefill] (-1.8691,0.499842)--(1.55681,0.0835205)--(1.55681,0.554759)--(-1.8691,0.971081)-- cycle;
\filldraw[conefill] (-1.8691,0.971081)--(1.55681,0.554759)--(1.55681,1.026)--(-1.8691,1.44232)-- cycle;
\filldraw[conefill] (-1.8691,1.44232)--(1.55681,1.026)--(1.55681,1.49724)--(-1.8691,1.91356)-- cycle;
\filldraw[conefill] (-1.8691,1.91356)--(1.55681,1.49724)--(1.55681,1.96848)--(-1.8691,2.3848)-- cycle;
\filldraw[conefill] (-1.8691,2.3848)--(1.55681,1.96848)--(1.55681,2.43971)--(-1.8691,2.85604)-- cycle;
\filldraw[conefill] (-1.8691,2.85604)--(1.55681,2.43971)--(1.55681,2.91095)--(-1.8691,3.32728)-- cycle;
\filldraw[conefill] (-1.8691,3.32728)--(1.55681,2.91095)--(1.55681,3.38219)--(-1.8691,3.79851)-- cycle;
\filldraw[conefill] (-1.8691,3.79851)--(1.55681,3.38219)--(1.55681,3.85343)--(-1.8691,4.26975)-- cycle;
\filldraw[conefill] (-1.8691,4.26975)--(1.55681,3.85343)--(1.55681,4.32467)--(-1.8691,4.74099)-- cycle;
\filldraw[conefill] (-1.8691,4.74099)--(1.55681,4.32467)--(1.55681,4.79591)--(-1.8691,5.21223)-- cycle;
\filldraw[conefill] (1.55681,-0.858957)--(4.25427,-0.595116)--(4.25427,-0.123877)--(1.55681,-0.387718)-- cycle;
\filldraw[conefill] (1.55681,-0.387718)--(4.25427,-0.123877)--(4.25427,0.347362)--(1.55681,0.0835205)-- cycle;
\filldraw[conefill] (1.55681,0.0835205)--(4.25427,0.347362)--(4.25427,0.818601)--(1.55681,0.554759)-- cycle;
\filldraw[conefill] (1.55681,0.554759)--(4.25427,0.818601)--(4.25427,1.28984)--(1.55681,1.026)-- cycle;
\filldraw[conefill] (1.55681,1.026)--(4.25427,1.28984)--(4.25427,1.76108)--(1.55681,1.49724)-- cycle;
\filldraw[conefill] (1.55681,1.49724)--(4.25427,1.76108)--(4.25427,2.23232)--(1.55681,1.96848)-- cycle;
\filldraw[conefill] (1.55681,1.96848)--(4.25427,2.23232)--(4.25427,2.70356)--(1.55681,2.43971)-- cycle;
\filldraw[conefill] (1.55681,2.43971)--(4.25427,2.70356)--(4.25427,3.1748)--(1.55681,2.91095)-- cycle;
\filldraw[conefill] (1.55681,2.91095)--(4.25427,3.1748)--(4.25427,3.64603)--(1.55681,3.38219)-- cycle;
\filldraw[conefill] (1.55681,3.38219)--(4.25427,3.64603)--(4.25427,4.11727)--(1.55681,3.85343)-- cycle;
\filldraw[conefill] (1.55681,3.85343)--(4.25427,4.11727)--(4.25427,4.58851)--(1.55681,4.32467)-- cycle;
\filldraw[conefill] (1.55681,4.32467)--(4.25427,4.58851)--(4.25427,5.05975)--(1.55681,4.79591)-- cycle;
\filldraw[conefill] (4.25427,-0.595116)--(4.96111,0.0784352)--(4.96111,0.549674)--(4.25427,-0.123877)-- cycle;
\filldraw[conefill] (4.25427,-0.123877)--(4.96111,0.549674)--(4.96111,1.02091)--(4.25427,0.347362)-- cycle;
\filldraw[conefill] (4.25427,0.347362)--(4.96111,1.02091)--(4.96111,1.49215)--(4.25427,0.818601)-- cycle;
\filldraw[conefill] (4.25427,0.818601)--(4.96111,1.49215)--(4.96111,1.96339)--(4.25427,1.28984)-- cycle;
\filldraw[conefill] (4.25427,1.28984)--(4.96111,1.96339)--(4.96111,2.43463)--(4.25427,1.76108)-- cycle;
\filldraw[conefill] (4.25427,1.76108)--(4.96111,2.43463)--(4.96111,2.90587)--(4.25427,2.23232)-- cycle;
\filldraw[conefill] (4.25427,2.23232)--(4.96111,2.90587)--(4.96111,3.37711)--(4.25427,2.70356)-- cycle;
\filldraw[conefill] (4.25427,2.70356)--(4.96111,3.37711)--(4.96111,3.84835)--(4.25427,3.1748)-- cycle;
\filldraw[conefill] (4.25427,3.1748)--(4.96111,3.84835)--(4.96111,4.31959)--(4.25427,3.64603)-- cycle;
\filldraw[conefill] (4.25427,3.64603)--(4.96111,4.31959)--(4.96111,4.79082)--(4.25427,4.11727)-- cycle;
\filldraw[conefill] (4.25427,4.11727)--(4.96111,4.79082)--(4.96111,5.26206)--(4.25427,4.58851)-- cycle;
\filldraw[conefill] (4.25427,4.58851)--(4.96111,5.26206)--(4.96111,5.7333)--(4.25427,5.05975)-- cycle;
\filldraw[conefill] (4.96111,0.0784352)--(3.34659,0.699536)--(3.34659,1.17077)--(4.96111,0.549674)-- cycle;
\filldraw[conefill] (4.96111,0.549674)--(3.34659,1.17077)--(3.34659,1.64201)--(4.96111,1.02091)-- cycle;
\filldraw[conefill] (4.96111,1.02091)--(3.34659,1.64201)--(3.34659,2.11325)--(4.96111,1.49215)-- cycle;
\filldraw[conefill] (4.96111,1.49215)--(3.34659,2.11325)--(3.34659,2.58449)--(4.96111,1.96339)-- cycle;
\filldraw[conefill] (4.96111,1.96339)--(3.34659,2.58449)--(3.34659,3.05573)--(4.96111,2.43463)-- cycle;
\filldraw[conefill] (4.96111,2.43463)--(3.34659,3.05573)--(3.34659,3.52697)--(4.96111,2.90587)-- cycle;
\filldraw[conefill] (4.96111,2.90587)--(3.34659,3.52697)--(3.34659,3.99821)--(4.96111,3.37711)-- cycle;
\filldraw[conefill] (4.96111,3.37711)--(3.34659,3.99821)--(3.34659,4.46945)--(4.96111,3.84835)-- cycle;
\filldraw[conefill] (4.96111,3.84835)--(3.34659,4.46945)--(3.34659,4.94069)--(4.96111,4.31959)-- cycle;
\filldraw[conefill] (4.96111,4.31959)--(3.34659,4.94069)--(3.34659,5.41193)--(4.96111,4.79082)-- cycle;
\filldraw[conefill] (4.96111,4.79082)--(3.34659,5.41193)--(3.34659,5.88316)--(4.96111,5.26206)-- cycle;
\filldraw[conefill] (4.96111,5.26206)--(3.34659,5.88316)--(3.34659,6.3544)--(4.96111,5.7333)-- cycle;
\filldraw[conefill] (-4.9034,3.35711)--(-4.42043,2.01976)--(-4.42043,2.491)--(-4.9034,3.82835)-- cycle;
\filldraw[conefill] (-4.9034,3.82835)--(-4.42043,2.491)--(-4.42043,2.96224)--(-4.9034,4.29959)-- cycle;
\filldraw[conefill] (-4.9034,4.29959)--(-4.42043,2.96224)--(-4.42043,3.43348)--(-4.9034,4.77083)-- cycle;
\filldraw[conefill] (-4.42043,-2.22139)--(-1.8691,-3.27007)--(-1.8691,-2.79883)--(-4.42043,-1.75015)-- cycle;
\filldraw[conefill] (-4.42043,-1.75015)--(-1.8691,-2.79883)--(-1.8691,-2.32759)--(-4.42043,-1.27891)-- cycle;
\filldraw[conefill] (-4.42043,-1.27891)--(-1.8691,-2.32759)--(-1.8691,-1.85635)--(-4.42043,-0.80767)-- cycle;
\filldraw[conefill] (-4.42043,-0.80767)--(-1.8691,-1.85635)--(-1.8691,-1.38511)--(-4.42043,-0.336431)-- cycle;
\filldraw[conefill] (-4.42043,-0.336431)--(-1.8691,-1.38511)--(-1.8691,-0.913874)--(-4.42043,0.134808)-- cycle;
\filldraw[conefill] (-4.42043,0.134808)--(-1.8691,-0.913874)--(-1.8691,-0.442635)--(-4.42043,0.606046)-- cycle;
\filldraw[conefill] (-4.42043,0.606046)--(-1.8691,-0.442635)--(-1.8691,0.0286035)--(-4.42043,1.07729)-- cycle;
\filldraw[conefill] (-4.42043,1.07729)--(-1.8691,0.0286035)--(-1.8691,0.499842)--(-4.42043,1.54852)-- cycle;
\filldraw[conefill] (-4.42043,1.54852)--(-1.8691,0.499842)--(-1.8691,0.971081)--(-4.42043,2.01976)-- cycle;
\filldraw[conefill] (-4.42043,2.01976)--(-1.8691,0.971081)--(-1.8691,1.44232)--(-4.42043,2.491)-- cycle;
\filldraw[conefill] (-4.42043,2.491)--(-1.8691,1.44232)--(-1.8691,1.91356)--(-4.42043,2.96224)-- cycle;
\filldraw[conefill] (-4.42043,2.96224)--(-1.8691,1.91356)--(-1.8691,2.3848)--(-4.42043,3.43348)-- cycle;
\filldraw[red]  (-1.8691,0.971081)  circle(2pt);
\draw[thick,red]  (-1.8691,0.499842) -- (-1.8691,0.971081) --(1.55681,0.554759)--(4.25427,0.818601) -- (4.25427,1.28984)--(4.96111,1.96339) -- (4.96111,2.43463);

\draw[thick,red] (-4.9034,2.41464)--(-4.42043,1.07729)--(-4.42043,1.54852)--(-1.8691,0.499842);
\filldraw[red]  (-1.8691,0.971081)  circle(2pt);
\filldraw[red]  (4.96111,2.43463)  circle(2pt);
\filldraw[red]  (-3.09201,3.56189)  circle(2pt);
\node at (-1.4691,0.71081) {${}_{w_n^0}$};
\node at  (5.4111,2.43463) {${}_{w_n^1}$};
\node at (-2.709201,3.86189) {${}_{w_n^2}$};
\end{tikzpicture}
\caption{A pictorial representation of the way in which the monodromy matrix wraps around to similar points for a (9,6)-reduction. The points $w_0^{p}$ given for reference to figure \ref{figredgeneral}. \label{loop}}
\end{figure}

To obtain a Lax representation for the system of ordinary difference equations, \eqref{autred}, we define two operators, $A_n$ and $B_n$, associated with the shifts $(l,m) \to (l+s_1,m+s_2)$ and the generating shift, $(l,m) \to (l+c,m+d)$, respectively. These operators have the effect
\begin{subequations}\label{Laxaut}
\begin{align}
\label{LaxLn}\phi_n &= A_n \phi_n,\\
\label{LaxMn}\phi_{n+1} &= B_n \phi_n,
\end{align}
\end{subequations}
where one representation\footnote{In practice, the product follows the path of a standard staircase \cite{vdKRQ:Staircase}.}, that is simple to write, is as follows:
\begin{align*}
A_n &\mapsfrom \prod_{j=0}^{s_2-1} M_{l+s_1,m+j}\prod_{i=0}^{s_1-1} L_{l+i,m},\\
B_n &\mapsfrom \prod_{j=0}^{d-1} M_{l+c,m+j}\prod_{i=0}^{c-1} L_{l+i,m},
\end{align*}
where the dependence on $n$ and $p$ is specified by 
\begin{align*}
L_{l,m} (w_{l,m},w_{l+1,m};\gamma) &\mapsto L_n^p(\gamma) = L_n^{p}(w_{n}^{p},w_{n-b}^{p+d}; \gamma),\\
M_{l,m} (w_{l,m},w_{l+1,m};\gamma) &\mapsto M_n^p(\gamma) = M_n^{p}(w_{n}^{p},w_{n+a}^{p-c}; \gamma).
\end{align*}
The compatibility condition,
\begin{equation}\label{Compaut}
A_{n+1}B_n - B_n A_n = 0,
\end{equation}
is equivalent to imposing \eqref{autred}. We call $A_n$ the monodromy matrix for the following reason: by identifying all points in $\mathbb{Z}^2$ that are multiples of $(s_1,s_2)$ apart, we may consider the space in which the new system exists as being cylindrical. We wrap around in a manner that connects the points that are identified by the periodic reduction. The monodromy matrix, rather than presenting a trivial action as \eqref{LaxLn} suggests, expresses the action of wrapping around the cylinder, as in figure \ref{loop}.

The monodromy matrix can be expressed as a function of the $s_1 + s_2$ initial conditions, $(w_n^0, w_{n+1}^0, \ldots, w_{a+b-1}^{g-1})$, by following the standard staircase. Geometrically, the standard staircase is the path between two lines which squeeze a set of squares with the same values, i.e., a set of squares shifted by $(s_1,s_2)$ \cite{VanderKamp:IVPs}.

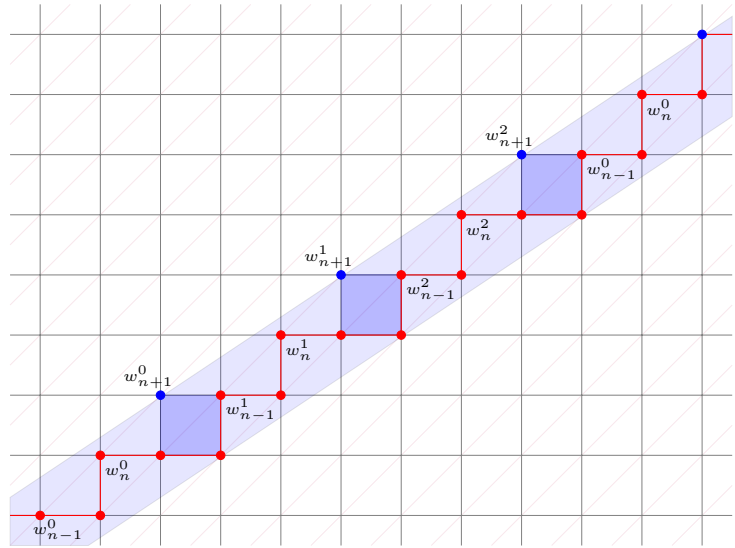
\begin{figure}[!ht]
\begin{tikzpicture}[scale=.8] 
  \draw[thin,purple,opacity=.1] (-1.5,6.5) -- (-.5,7.5);
  \draw[thin,purple,opacity=.1] (-1.5,5.5) -- (.5,7.5);
  \draw[thin,purple,opacity=.1] (-1.5,4.5) -- (1.5,7.5);
  \draw[thin,purple,opacity=.1] (-1.5,3.5) -- (2.5,7.5);
  \draw[thin,purple,opacity=.1] (-1.5,2.5) -- (3.5,7.5);
  \draw[thin,purple,opacity=.1] (-1.5,1.5) -- (4.5,7.5);
  \draw[thin,purple,opacity=.1] (-1.5,.5) -- (5.5,7.5);
  \draw[thin,purple,opacity=.1] (-1.5,-.5) -- (6.5,7.5);
  \draw[thin,purple,opacity=.1] (-1.5,-1.5) -- (7.5,7.5);
  \draw[thin,purple,opacity=.1] (-.5,-1.5) -- (8.5,7.5);
  \draw[thin,purple,opacity=.1] (.5,-1.5) -- (9.5,7.5);
  \draw[thin,purple,opacity=.1] (1.5,-1.5) -- (10.5,7.5);
  \draw[thin,purple,opacity=.1] (2.5,-1.5) -- (10.5,6.5);
  \draw[thin,purple,opacity=.1] (3.5,-1.5) -- (10.5,5.5);
  \draw[thin,purple,opacity=.1] (4.5,-1.5) -- (10.5,4.5);
  \draw[thin,purple,opacity=.1] (5.5,-1.5) -- (10.5,3.5);
  \draw[thin,purple,opacity=.1] (6.5,-1.5) -- (10.5,2.5);
  \draw[thin,purple,opacity=.1] (7.5,-1.5) -- (10.5,1.5);
  \draw[thin,purple,opacity=.1] (8.5,-1.5) -- (10.5,.5);
  \draw[thin,purple,opacity=.1] (9.5,-1.5) -- (10.5,-.5);
 \draw[very thin,color=gray] (-1.5,-1.5) grid (10.5,7.5);
 \draw[fill=blue,opacity=.2] (1,1)-- (2,1) -- (2,0)-- (1,0) -- cycle;
 \draw[fill=blue,opacity=.2] (7,5)-- (8,5) -- (8,4)-- (7,4) -- cycle;
 \draw[fill=blue,opacity=.2] (4,3)-- (5,3) -- (5,2)-- (4,2) -- cycle;
 \draw[fill=blue,opacity=.1] (-.2,-1.5)-- (-1.5,-1.5) -- (-1.5,-.7) -- (10.5,7.3) -- (10.5,5.6333)  -- cycle;
  \draw[red] (-1.5,-1) -- (-1,-1) -- (0,-1) -- (0,0) -- (1,0)-- (2,0) -- (2,1)--(3,1)--(3,2)--(4,2)--(5,2)--(5,3)--(6,3)--(6,4)--(7,4)--(8,4)--(8,5)--(9,5) --(9,6) -- (10,6) -- (10,7) -- (10.5,7);
  \filldraw[red] (-1,-1) circle(2pt);
  \filldraw[red] (0,-1) circle(2pt);
  \filldraw[red] (0,0) circle(2pt);
  \filldraw[red] (1,0) circle(2pt);
  \filldraw[red] (2,0) circle(2pt);
  \filldraw[red] (2,1) circle(2pt);
  \filldraw[red] (3,1) circle(2pt);
  \filldraw[red] (3,2) circle(2pt);
  \filldraw[red] (4,2) circle(2pt);
  \filldraw[red] (5,2) circle(2pt);
  \filldraw[red] (5,3) circle(2pt);
  \filldraw[red] (6,3) circle(2pt);
  \filldraw[red] (6,4) circle(2pt);
  \filldraw[red] (7,4) circle(2pt);
  \filldraw[red] (8,4) circle(2pt);
  \filldraw[red] (8,5) circle(2pt);
  \filldraw[red] (9,5) circle(2pt);
  \filldraw[red] (9,6) circle(2pt);
  \filldraw[red] (10,6) circle(2pt);
 \begin{scope}[xshift = .3cm,yshift=-.25cm]
    \node at (-1,-1) {\tiny$w_{n-1}^0$};
    \node at (0,0) {\tiny$w_n^0$};
    \node at (2.2,1) {\tiny$w_{n-1}^1$};
    \node at (3,2) {\tiny$w_n^1$};
    \node at (5.2,3) {\tiny$w_{n-1}^2$};
    \node at (6,4) {\tiny$w_n^2$};
    \node at (8.2,5) {\tiny$w_{n-1}^0$};
    \node at (9,6) {\tiny$w_n^0$};
  \end{scope}
 \begin{scope}[xshift = -.2cm,yshift=.3cm]
    \node at (1,1) {\tiny{$w_{n+1}^0$}};
    \node at (4,3) {\tiny$w_{n+1}^1$};
    \node at (7,5) {\tiny$w_{n+1}^2$};
  \end{scope}
  \filldraw[blue] (1,1) circle(2pt);
  \filldraw[blue] (4,3) circle(2pt);
  \filldraw[blue] (7,5) circle(2pt);
  \filldraw[blue] (10,7) circle(2pt);
\end{tikzpicture}
\caption{The full labelling of variables in the $(9,6)$-reduction of figure \ref{loop}. In this example, the shift $(p,n) \to (p+1,n)$ corresponds to the shift $(a,b) = (3,2)$ and the shift $(p,n) \to (p,n+1)$ corresponds to the shift $(c,d) = (1,1)$. \label{figredgeneral}}
\end{figure}

One advantage of the generating shift is that every other shift in $n$ may be expressed as some power of the generating shift by construction \cite{vdKRQ:Staircase}. Furthermore, this generating shift allows us to constrain the non-linear component, where we need to use \eqref{autred}, to just $g$ places. We have illustrated the standard staircase and generating shift in figure \ref{figredgeneral}.

In the example defined by figure \ref{figredgeneral}, if we allow our monodromy matrix to follow the standard staircase, the monodromy matrix is
\begin{align*}
A_n \mapsfrom& L_{l+8,m+6}M_{l+8,m+5}L_{l+7,m+5}L_{l+6,m+5}M_{l+6,m+4}L_{l+5,m+4}M_{l+5,m+3}\\
&L_{l+4,m+3}L_{l+3,m+3}M_{l+3,m+2}L_{l+2,m+2}M_{l+2,m+1}L_{l+1,m+1}L_{l,m+1}M_{l,m},
\end{align*}
and the other half of the Lax pair is
\begin{align*}
B_n \mapsfrom M_{l+1,m}L_{l,m}.
\end{align*}
The resulting compatibility condition, \eqref{Compaut}, gives the evolution equations for the $w_{n+1}^i$, $i = 0,1,2$:
\begin{align*}
&Q(w_{n-2}^1, w_{n-4}^2, w_{n+1}^0, w_{n-1}^1; \alpha,\beta) = 0, \\
&Q(w_{n-2}^2, w_{n-4}^0, w_{n+1}^1, w_{n-1}^2; \alpha,\beta) = 0, \\
&Q(w_{n-2}^0, w_{n-4}^1, w_{n+1}^2, w_{n-1}^0; \alpha,\beta) = 0. 
\end{align*}
In general, this procedure gives us an $s_1 + s_2$ dimensional mapping, 
\[
\phi : \mathbb{C}^{s_1+s_2} \to \mathbb{C}^{s_1+s_2},
\]
which, applied to $( w_{n}^0,  w_{n+1}^0, \ldots ,w_{n+a+b-1}^{g-1})$, gives $( w_{n+1}^0,  w_{n+2}^0, \ldots ,w_{n+a+b}^{g-1})$. This new set of values forms a new standard staircase. As a matter of fact, this new standard staircase is the old one translated by the generating shift.

\subsection{Nonautonomous reductions}

To deautonomize this theory, we consider the $\alpha$ and $\beta$ to be functions of $l$ and $m$. As $L_{l,m}$ and $M_{l,m}$ are shifted in only $m$ and $l$ respectively in the compatibility condition, \eqref{comp}, replacing $\alpha$ and $\beta$ with $\alpha_l$ and $\beta_m$, which are arbitrary functions of $l$ and $m$ respectively, preserves the Lax integrability. Hence, our basic non-autonomous lattice equations may be considered to be of the form
\begin{equation}\label{non-autQ}
Q(w_{l,m},w_{l+1,m},w_{l,m+1},w_{l+1,m+1};\alpha_l,\beta_m) = 0,
\end{equation}
where the Lax representation is specified by \eqref{LaxForm} where
\begin{subequations}\label{nonautLaxmats}
\begin{align}
L_{l,m} &= \left.\lambda_{l,m} \begin{pmatrix} -\dfrac{\partial Q(x,u,v,0;\alpha_l,\gamma )}{\partial v} & -Q(x,u,0,0;\alpha_l,\gamma ) \\
 \dfrac{\partial^2 Q(x,u,v,y;\alpha_l,\gamma )}{\partial v \partial y} & \dfrac{\partial Q(x,u,0,y;\alpha_l ,\gamma )}{\partial y}\end{pmatrix}\right|_{\begin{array}{c} x= w_{l,m}\\ u = w_{l+1,m}\end{array}},\\
M_{l,m} &= \left. \mu_{l,m} \begin{pmatrix} -\dfrac{\partial Q(x,u,v,0;\beta_m,\gamma )}{\partial u} & -Q(x,0,v,0;\beta_m,\gamma ) \\
 \dfrac{\partial^2 Q(x,u,v,y;\beta_m,\gamma )}{\partial u \partial y} & \dfrac{\partial Q(x,0,v,y;\beta_m ,\gamma )}{\partial y}\end{pmatrix} \right|_{\begin{array}{c} x= w_{l,m}\\ v = w_{l,m+1}\end{array}},
\end{align}
\end{subequations}
where $\gamma$ is a spectral parameter and the prefactors, $\lambda_{l,m}$ and $\mu_{l,m}$, are chosen to satisfy the compatibility conditions, in an analogous manner to the autonomous case. 

If one assumes that the $\alpha$ and $\beta$ are functions of both $l$ and $m$, i.e., $\alpha = \alpha_{l,m}$ and $\beta = \beta_{l,m}$, then demanding that $\alpha_{l,m}$ is independent of $m$ and $\beta_{l,m}$ is independent of $l$ has also been shown to be a necessary condition for singularity confinement for equations in the ABS list \cite{Gramani:ABSSC}. The above constitutes a Lax pair interpretation of this constraint. 

Let us now specialise our choice of systems to those that admit representations of the additive form
\begin{equation}\label{q:add}
Q(w_{l,m},w_{l+1,m},w_{l,m+1},w_{l+1,m+1};\alpha_l - \beta_m) = 0,
\end{equation}
or the multiplicative form 
\begin{equation}\label{q:mult}
Q\left(w_{l,m},w_{l+1,m},w_{l,m+1},w_{l+1,m+1}; \dfrac{\alpha_l}{\beta_m}\right) = 0,
\end{equation}
with a possible additional dependence on $\alpha_{l+1} - \alpha_l$ and $\beta_{m+1} - \beta_m$ in the additive case, or $\alpha_{l+1}/\alpha_l$ and $\beta_{m+1}/\beta_m$ in the multiplicative case. A list of transformed equations appears in table \ref{tablenotautoQlist}, where subscripts $m$ and $a$ denote those functions, \eqref{Q}, dependent on a multiplicative or additive combination of $\alpha_l$ and $\beta_m$ respectively. This list is restricted to equations that we could find transformations to forms admitting additive or multiplicative reductions. This does not include the equations Q3 or Q4.

\begin{table}[!ht]
\begin{tabular}{l l} \toprule
ABS & $Q(x,u,v,y;\alpha_l,\beta_m)$ \\ \toprule
$\mathrm{H1}_a$ & $(w_{l,m}-w_{l+1,m+1})(w_{l+1,m}-w_{l,m+1}) + \beta_m-\alpha_l$\\ \midrule
$\mathrm{H1}_m$ & $\left(w_{l,m}-\dfrac{\beta_{m+1}}{\beta_m}w_{l+1,m+1}\right)\left(w_{l+1,m}-\dfrac{\beta_{m+1}}{\beta_m}w_{l,m+1}\right) + 1-\dfrac{\alpha_l^2}{\beta_m^2}$\\ \midrule
$\mathrm{H2}_m$ & $\left(w_{l,m}-\dfrac{\beta_{m+1}}{\beta_m}w_{l+1,m+1}\right)\left(w_{l+1,m}-\dfrac{\beta_{m+1}}{\beta_m}w_{l,m+1}\right)-\dfrac{\alpha_l^2}{\beta_m^2}$ \\
& $+ \left(1-\dfrac{\alpha_l}{\beta_m}\right)\left(w_{l,m}+w_{l+1,m}+\dfrac{\beta_{m+1}}{\beta_m}\left(w_{l,m+1}+w_{l+1,m+1}\right)\right) + 1$\\ \midrule 
$\mathrm{H3}_m^{\delta=0}$ & $\dfrac{\alpha_l}{\beta_m}(w_{l,m}w_{l+1,m}+w_{l,m+1}w_{l+1,m+1}) -(w_{l,m}w_{l,m+1}+w_{l+1,m}w_{l+1,m+1})$\\ \midrule
$\mathrm{H3}_m^{\delta \neq 0}$ & $\dfrac{\alpha_l^2}{\beta_m^2}\left(w_{l,m}w_{l+1,m}+\dfrac{\beta_{m+1}^2}{\beta_m^2}w_{l,m+1}w_{l+1,m+1}\right)$\\
& $-\dfrac{\beta_{m+1}}{\beta_m}\left(w_{l,m}w_{l,m+1}+w_{l+1,m}w_{l+1,m+1}\right) + \delta \left(\dfrac{\alpha_l^4}{\beta_m^4}-1\right)$ \\ \midrule
$\mathrm{Q1}_m^{\delta= 0}$ & $\dfrac{\alpha_l}{\beta_m}(w_{l,m}-w_{l,m+1})(w_{l+1,m}-w_{l+1,m+1})$  \\
& $- (w_{l,m}-w_{l+1,m})(w_{l,m+1}-w_{l+1,m+1})$\\ \midrule
$\mathrm{Q1}_m^{\delta \neq 0}$ & $\dfrac{\alpha_l^2}{\beta_m^2}\left(w_{l,m}-\dfrac{\beta_{m+1}}{\beta_m}w_{l,m+1}\right)\left(w_{l+1,m}-\dfrac{\beta_{m+1}}{\beta_m}w_{l+1,m+1}\right)$ \\
& $ -\dfrac{\beta_{m+1}}{\beta_m}\left(w_{l,m} -w_{l+1,m}\right)\left(w_{l,m+1}-w_{l+1,m+1}\right) + \dfrac{\delta\alpha_l^2}{\beta_m^2}\left(\dfrac{\alpha_l^2}{\beta_m^2}-1\right)$\\ \midrule
$\mathrm{Q2}_m$ & $\dfrac{\alpha _l}{\beta_m} \left(w_{l+1,m}-\dfrac{\beta_{m+1}^2}{\beta_m^2}w_{l+1,m+1}\right) \left(w_{l,m}-\dfrac{\beta_{m+1}^2}{\beta_m^2}w_{l,m+1}\right)$ \\ & $-\dfrac{\beta_{m+1}^2}{\beta_m^2}\left(w_{l,m}-w_{l+1,m}\right) \left(w_{l,m+1}-w_{l+1,m+1}\right)$ \\ 
& $-\dfrac{\alpha _l}{\beta_m} \left(\dfrac{\alpha _l}{\beta_m}-1\right) \left(w_{l,m}+w_{l+1,m}+\dfrac{\beta_{m+1}^2}{\beta_m^2}w_{l,m+1}+\dfrac{\beta_{m+1}^2}{\beta_m^2}w_{l+1,m+1}\right)$ \\ 
& $- \dfrac{\alpha _l}{\beta _m} \left(\dfrac{\alpha _l}{\beta _m}-1\right) \left(\dfrac{\alpha _l^2}{\beta_m^2}-\dfrac{\alpha _l}{\beta _m}+1\right)$ 
\\ \bottomrule
\end{tabular}
\caption{A list of various lattice equations (taken from \cite{ABS:ListI,ABS:ListII}) in a suitable form for non-autonomous reductions.}
\label{tablenotautoQlist}
\end{table}

With each lattice equation written in terms of $\alpha_l - \beta_m$ or $\alpha_l/\beta_m$, the necessary requirement for \eqref{periodicity} to be consistent is the requirement that 
\begin{align*}
\alpha_l - \beta_m = \alpha_{l+s_1} - \beta_{m+s_2} ,
\end{align*}
in the additive case, and
\begin{align*}
\dfrac{\alpha_l}{\beta_m} = \dfrac{\alpha_{l+s_1}}{\beta_{m+s_2}},
\end{align*}
in the multiplicative case. By a separation of variables argument, we define $h$ and $q$ by letting
\begin{subequations}
\begin{align}
\label{hseq}\alpha_{l+s_1} - \alpha_l = \beta_{m+s_2} - \beta_{m} &:= habg,\\
\label{qseq}\dfrac{\alpha_{l+s_1}}{\alpha_l} = \dfrac{\beta_{m+s_2}}{\beta_{m}} &:= q^{abg},
\end{align}
\end{subequations}
in the additive and multiplicative cases respectively. Although it is not a technical requirement, we will assume that $h$ is not $0$ and that $q$ is not a root of unity. We solve the additive and multiplicative case by letting
\begin{align*}
\alpha_l = hlb + a_l, && \beta_m = hma + b_m,\\
\alpha_l = a_lq^{bl}, && \beta_m = b_mq^{am},
\end{align*}
where $a_l$ and $b_m$ are sequences that are periodic of order $s_1$ and $s_2$ respectively (not related to the constants, $a$ and $b$). This choice of $\alpha_l$ and $\beta_m$ ensures the consistency of the reduction with as many degrees of freedom as the sum of the orders of the difference equations satisfied by $\alpha_l$ and $\beta_m$, \eqref{hseq} and \eqref{qseq}, i.e., $s_1 + s_2$.

To provide a non-autonomous Lax pair for the non-autonomous reduction, we need to choose a spectral variable, $x$, in a manner that couples a linearly independent direction with the spectral variable, $\gamma$. While any linearly independent direction may be considered a valid choice, we present a simple choice. Our choice of spectral parameter is specified by introducing the variable $k = l$ and $x = hbk - \gamma$ in the additive case and $x = q^{bk}/\gamma$ in the multiplicative case. In the additive case
\begin{align*}
L_{l,m} &= L_{l,m}(\alpha_l-\gamma) \mapsto L_n(a_l + x),\\
M_{l,m} &= M_{l,m}((\beta_m - \alpha_l) + (\alpha_l-\gamma)) \mapsto M_n(x + hn +b_m),
\end{align*}
and in the multiplicative case
\begin{align*}
L_{l,m} &= L_{l,m}(\alpha_l/\gamma) = L_{l,m}(a_lx),\\
M_{l,m} &= M_{l,m}((\beta_m/\alpha_l)(\alpha_l/\gamma)) = M_{l,m}(b_mxq^n).
\end{align*}
This gives us a non-standard Lax pair, which, in the additive case reads
\begin{align*}
Y_n(x+abgh) &= A_n(x)Y_n(x),\\
Y_{n+1}(x+cbh) &= B_n(x)Y_n(x),
\end{align*}
and in the multiplicative case reads
\begin{align*}
Y_n(q^{abg}x) &= A_n(x)Y_n(x),\\
Y_{n+1}(q^{cb}x) &= B_n(x)Y_n(x),
\end{align*}
where
\begin{subequations}\label{prods}
\begin{align}
\label{ProdA}A_n(x) \mapsfrom & \prod_{j=0}^{s_2-1} M_{l+s_1,m+j}\prod_{i=0}^{s_1-1} L_{l+i,m},\\
\label{ProdB}B_n(x) \mapsfrom& \prod_{j=0}^{d-1} M_{l+c,m+j}\prod_{i=0}^{c-1} L_{l+i,m}.
\end{align}
\end{subequations}
The compatibility conditions,
\begin{align*}
A_{n+1}(x+cbh)B_n(x) &= B_n(x+abgh)A_n(x),\\
A_{n+1}(q^{cb}x)B_n(x) &= B_n(q^{abg}x)A_n(x),
\end{align*}
in the additive and multiplicative cases respectively, gives us \eqref{nautred}. This choice of spectral variable has the advantage that the spectral matrix and deformation matrix, $A_n(x)$ and $B_n(x)$, have a simple dependence on the independent variable, $n$.

\section{Some simple examples}

In this section, we present some examples of the theory above. An example that has appeared recently is the example of $q$-$\mathrm{P}_{VI}$ as a reduction of the discrete modified Korteweg-de Vries equation \cite{Ormerod:qP6}, which also gave rise, via ultradiscretization, to the first known Lax representation of $u$-$\mathrm{P}_{VI}$.

\subsection{Autonomous example}

We consider some additive examples, in particular, we will consider reductions of the discrete potential Korteweg-de Vries equation,
\begin{equation}\label{H1}
(w_{l,m}-w_{l+1,m+1})(w_{l+1,m}-w_{l,m+1}) = \alpha - \beta,
\end{equation}
labelled as $H1_a$ in table \ref{tablenotautoQlist}, which possesses a Lax representation of the form \eqref{LaxForm} where $L_{l,m}$ and $M_{l,m}$ are specified by
\begin{subequations}
\begin{align}
L_{l,m} =& \begin{pmatrix}
      w_{l,m} & \alpha -\gamma-w_{l,m} w_{l+1,m}\\
      1       & -w_{l+1,m}
    \end{pmatrix},\\
M_{l,m} =&
    \begin{pmatrix}
      w_{l,m} & \beta -\gamma-w_{l,m} w_{l,m+1} \\
      1       & -w_{l,m+1}    \end{pmatrix}.
\end{align}
\end{subequations}
Let us consider a reduction, \eqref{periodicity}, where $s_1 = 2$ and $s_2= 1$, with a labelling indicated in figure \ref{simpex13label}. This gives us $g=1$, $a=2$ and $b=1$, hence $n= 2m - l$, and the direction that characterises the generating shift, $(c,d) $, is chosen to be $(1,1)$.

\begin{figure}[!ht]
\begin{tikzpicture}[scale=.7] 
 \draw[very thin,color=gray] (-1.5,-1.5) grid (3.5,2.5);
    \draw[red] (-1.5,-1) -- (0,-1) -- (0,0) -- (1,0) -- (2,0) -- (2,1) -- (3,1)-- (3.5,1);
    \begin{scope}[yshift=1cm]
    \draw[red] (-1.5,-1) -- (0,-1) -- (0,0) -- (1,0) -- (2,0) -- (2,1) -- (3,1)-- (3.5,1);
    \end{scope}
    \filldraw[red] (0,-1) circle(2pt);
    \filldraw[red] (-1,-1) circle(2pt);
    \filldraw[red] (0,0) circle(2pt);
    \filldraw[red] (1,0) circle(2pt);
    \filldraw[red] (2,0) circle(2pt);
    \filldraw[red] (2,1) circle(2pt);
    \filldraw[red] (3,1) circle(2pt);
    \filldraw[blue] (-1,0) circle(2pt);
    \filldraw[blue] (0,1) circle(2pt);
    \filldraw[blue] (1,1) circle(2pt);
    \filldraw[blue] (2,2) circle(2pt);
    \filldraw[blue] (3,2) circle(2pt);
    \begin{scope}[xshift = .3cm,yshift=-.3cm]
	  \node at (-1,-1) {$w_1$};
	  \node at (0,-1) {$w_0$};
    \node at (0,0) {$w_2$};
    \node at (1,0) {$w_1$};
    \node at (2,0) {$w_0$};
    \node at (2,1) {$w_2$};
    \node at (3,1) {$w_1$};
    \node at (1,1) {$w_3$};
    \node at (0,1) {$w_4$};
    \node at (2,2) {$w_4$};
    \node at (3,2) {$w_3$};
    \node at (-1,0) {$w_3$};
    \end{scope}
\end{tikzpicture}
\caption{The labelling of initial conditions with (2,1) periodicity and an evolution in the $(1,1)$-direction.\label{simpex13label}}
\end{figure}

The product formula for the monodromy matrix, $A_n$, and the matrix that is related to the generating shift, $B_n$, are
\begin{align*}
A_n =& M_{l+2,m} L_{l+1,m} L_{l,m}\\
&= \left(
\begin{array}{cc}
 w_n & \beta -\gamma -w_n w_{n+2} \\
 1 & -w_{n+2}
\end{array}
\right)\left(
\begin{array}{cc}
 w_{n+1} & \alpha -\gamma -w_n w_{n+1} \\
 1 & -w_n
\end{array}
\right)\\
& \left(
\begin{array}{cc}
 w_{n+2} & \alpha -\gamma -w_{n+1} w_{n+2} \\
 1 & -w_{n+1}
\end{array}
\right),\\
B_n =& M_{l+1,m}L_{l,m}\\
&= \left(
\begin{array}{cc}
 w_{n+1} & \beta -\gamma -w_{n+1} w_{n+3} \\
 1 & -w_{n+3}
\end{array}
\right)\left(
\begin{array}{cc}
 w_{n+2} & \alpha -\gamma -w_{n+1} w_{n+2} \\
 1 & -w_{n+1}
\end{array}
\right).
\end{align*}
The compatibility condition, given by \eqref{Compaut}, reads
\begin{align*}
M_{l+3,m+1}L_{l+2,m+1}&L_{l+1,m+1}M_{l+1,m}L_{l,m} \\
&= M_{l+3,m+1}L_{l+2,m+1}M_{l+2,m}L_{l+1,m}L_{l,m}.
\end{align*}
This simplifies to
\begin{align*}
&L_{l+1,m+1}M_{l+1,m} = M_{l+2,m}L_{l+1,m},\\
&\left(
\begin{array}{cc}
 w_{n+3} & \alpha -\gamma -w_{n+2} w_{n+3} \\
 1 & -w_{n+2}
\end{array}
\right)\left(
\begin{array}{cc}
 w_{n+1} & \beta -\gamma -w_{n+1} w_{n+3} \\
 1 & -w_{n+3}
\end{array}
\right)\\ &=
\left(
\begin{array}{cc}
 w_n & \beta -\gamma -w_n w_{n+2} \\
 1 & -w_{n+2}
\end{array}
\right)\left(
\begin{array}{cc}
 w_{n+1} & \alpha -\gamma -w_n w_{n+1} \\
 1 & -w_n
\end{array}
\right),
\end{align*}
which defines the evolution of this autonomous reduction to be given by the equation
\begin{equation}\label{H1autred}
(w_{n}- w_{n+3})(w_{n+1}-w_{n+2}) = \alpha - \beta.
\end{equation}
If we let $y_n = w_{n} - w_{n+1}$, this equation is equivalent to
\begin{equation}\label{QRT}
y_{n-1} + y_n + y_{n+1} = \dfrac{\alpha - \beta}{y_n},
\end{equation}
which is a well known example of a second order difference equation of QRT type.

\subsection{Nonautonomous example}

The autonomous equation and Lax representation generalise naturally to the non-autonomous case by replacing $\alpha$ and $\beta$ by $\alpha_l$ and $\beta_m$ respectively. Furthermore, we may satisfy the periodicity constraint, 
\[
\alpha_{l+2} - \alpha_l = \beta_{m+1} - \beta_m := 2h.
\]
We solve this constraint by letting
\[
\alpha_l = h l + a_l, \hspace{1cm} \beta_m = 2hm + b_m,
\]
where $a_l$ is periodic of order two and $b_m$ is constant, and hence, may be taken to be $0$ without loss of generality. The evolution equation for this system may be represented as an application of the nonautonomous version of \eqref{H1} translated by the vector $(1,1)$;
\begin{equation}\label{nonautoSimpred}
(w_{n}- w_{n+3})(w_{n+1}-w_{n+2}) = \alpha_{l+1} - \beta_{m+1}.
\end{equation}
recalling that $n = 2m-l$. The increment in $l$ and $m$ by $1$ directly corresponds to the increment in $n$ by $1$. In the simplest case where $a_l = a_1$ is constant (rather than periodic), letting
\[
y_n = w_n - w_{n+1},
\]
results in the evolution equation
\[
y_n + y_{n+1} + y_{n+2} = \dfrac{\alpha_{l+1} - \beta_{m+1}}{y_{n+1}}= \dfrac{-hn-h + a_1}{y_{n+1}},
\]
or alternatively 
\begin{equation}\label{dPI}
y_{n-1} + y_n + y_{n+1} = \dfrac{-hn + a_1}{y_n}.
\end{equation} 

To form the Lax pair for this reduction, we choose a spectral variable to be
\[
x = h l - \gamma.
\]
Since $\alpha_l$ and $\beta_m$ appear with $\gamma$ in $L_{l,m}$ and $M_{l,m}$, this gives us
\begin{align*}
\alpha_l - \gamma &= hl - \gamma + a_1 = a_1 + x,\\
\beta_m - \gamma &= 2hm - \gamma = hn + x.
\end{align*}
which means we may interpret our $l$ and $m$ variables in terms of $x$ and $n$ variables. The $(l,m) \to (l+2,m+1)$ shift and $(l,m) \to (l+1,m+1)$ gives us $(n,x) \to (n,x+2h)$ and $(n,x) \to (n+1,x+h)$, hence, we have a linear system of the form
\begin{align*}
Y_n(x+2h) &= A_n(x)Y_n(x),\\
Y_{n+1}(x+h) &= B_n(x)Y_n(x),
\end{align*}
where
\begin{align*}
A_n(x) \mapsfrom& M_{l+2,m}L_{l+1,m}L_{l,m} \\
=& \left(
\begin{array}{cc}
 w_n & h n+x-w_n w_{n+2} \\
 1 & -w_{n+2}
\end{array}
\right)\left(
\begin{array}{cc}
 w_{n+1} & a_2 + x-w_n w_{n+1} \\
 1 & -w_n
\end{array}
\right)\\
&\left(
\begin{array}{cc}
 w_{n+2} & a_1+x-w_{n+1} w_{n+2} \\
 1 & -w_{n+1}
\end{array}
\right),\\
B_n(x) \mapsfrom & M_{l+1,m}L_{l,m},\\
=& \left(
\begin{array}{cc}
 w_{n+1} & h n+x-w_{n+1} w_{n+3} \\
 1 & -w_{n+3}
\end{array}
\right)\left(
\begin{array}{cc}
 w_{n+2} & a_1+x-w_{n+1} w_{n+2} \\
 1 & -w_{n+1}
\end{array}
\right).
\end{align*}
The compatibility condition is
\begin{equation}\label{compsimp}
A_{n+1}(x+h)B_n(x) = B_n(x+2h)A_n(x),
\end{equation}
which gives \eqref{nonautoSimpred}. To express everything in terms of $y_n = w_n - w_{n+1}$, we use the matrices obtained from applying a gauge transformation, $L_{l,m} \to S_{l+1,m}^{-1}L_{l,m}S_{l,m}$ and $M_{l,m} \to S_{l,m+1}^{-1}M_{l,m}S_{l,m}$, where
\[
S_{l,m} = \begin{pmatrix} 1 & w_{l,m} \\ 0 & 1 \end{pmatrix},
\]
in which case $A_n(x)$ and $B_n(x)$ are given explicitly in terms of products of matrices with entries expressible in terms of the $y_n$ variables:
\begin{align*}
A_n(x) &= \left(
\begin{array}{cc}
 -y_n-y_{n+1} & \left(y_n+y_{n+1}\right)^2+h n+x \\
 1 & -y_n-y_{n+1}
\end{array}
\right)\left(
\begin{array}{cc}
 y_n & a_1+h+y_n^2+x \\
 1 & y_n
\end{array}
\right)\\ 
&\left(
\begin{array}{cc}
 y_{n+1} & a_1+y_{n+1}^2+x \\
 1 & y_{n+1}
\end{array}
\right),\\
B_n(x) =& \left(
\begin{array}{cc}
 -y_{n+1}-y_{n+2} & \left(y_{n+1}+y_{n+2}\right)^2+h n+x \\
 1 & -y_{n+1}-y_{n+2}
\end{array}
\right)\\
&\left(
\begin{array}{cc}
 y_{n+1} & a_1+y_{n+1}^2+x \\
 1 & y_{n+1}
\end{array}
\right).
\end{align*}
for which \eqref{compsimp} now gives \eqref{dPI} as required. We recover Lax matrices for \eqref{QRT} by letting
\begin{align*}
A_n := \lim_{h \to 0} A_n(x),\\
B_n := \lim_{h \to 0} B_n(x),
\end{align*}
whose compatibility, \eqref{Compaut}, gives \eqref{QRT} where $\alpha -\beta = a_1$. This is a way in which the autonomous and nonautonomous reductions are related.

While the simple case above demonstrates the basic mechanisms in the method, the periodicity constraint allows us to build in an extra variable. If we allow the full generality of a periodic value of $a_l$, by letting 
\[
\alpha_l = \left\{ \begin{array}{c p{4cm}} hl + a_1 & where $l$ is odd, \\ hl + a_2 & where $l$ is even,
\end{array}\right.
\]
we obtain
\begin{align*}
A_n(x) \mapsfrom& M_{l+2,m}L_{l+1,m}L_{l,m} \\
A_n(x)=& \left(
\begin{array}{cc}
 w_n & h n+x-w_n w_{n+2} \\
 1 & -w_{n+2}
\end{array}
\right)\left(
\begin{array}{cc}
 w_{n+1} & a_2 + x-w_n w_{n+1} \\
 1 & -w_n
\end{array}
\right)\\
&\left(
\begin{array}{cc}
 w_{n+2} & a_1+x+w_{n+1} w_{n+2} \\
 1 & -w_{n+1}
\end{array}
\right),\\
B_n(x) \mapsfrom & M_{l+1,m}L_{l,m},\\
B_n(x)=& \left(
\begin{array}{cc}
 w_{n+1} & h n+x-w_{n+1} w_{n+3} \\
 1 & -w_{n+3}
\end{array}
\right)\left(
\begin{array}{cc}
 w_{n+2} & a_1+x-w_{n+1} w_{n+2} \\
 1 & -w_{n+1}
\end{array}
\right).
\end{align*}
The compatibility condition needs to take into account that the shift, $(n,x) \to (n+1,x+h)$, also shifts the position on the lattice, hence, we need to couple the non-linear component with a swapping of the roles of $a_1$ and $a_2$, hence, the compatibility condition, given by \eqref{compsimp}, results in the evolution
\begin{align*}
(w_n - w_{n+3})(w_{n+1}-w_{n+2}) = hn+a_2,\\
a_1 \to a_2 +h, \hspace{.7cm} a_2 \to a_1 - h. 
\end{align*}
If we restrict our attention to even powers of this map, the $a_i$ are constant. We proceed to specify a change of variables, which is motivated by some historical context. In the 1980's, Novikov and Veselov formalized the derivation of the Hamiltonian structure of hierarchies of soliton equations from their Lax representations \cite{NovikovVeselov:Hamiltonianvariables, KricheverPhoung}. The Darboux coordinates are the poles of the Baker-Akhiezer function and the eigenvalues of the spectral matrix at these poles, which for $2\times 2$ matrices coincides with the roots in the spectral variable of the off-diagonal elements and the diagonal elements evaluated at those roots. The Hamiltonian description of the isomonodromic deformations of \cite{Jimbo:Monodromy2} are expressed in these coordinates. While the link between the symplectic structure for discrete Painlev\'e equations and the discrete isomonodromic deformations is not as well developed, it is interesting to note that the parameterizations of many known discrete Lax pairs are provided in terms of these coordinates \cite{Murata2009, SakaiE6, Sakai:qP6}. Perhaps a discrete analogue of the Baker-Akhiezer function holds the key to linking the geometric theory and the theory of discrete isomonodromic deformations. It may also provide an algorithmic manner of describing the symplectic structure for hierarchies of equations.

Our new coordinates are $(y_n,z_n)$ where $y_n$ is the root in the spectral variable of the $(2,1)$-element and $z_n$ is a variable which parameterizes the diagonal elements at this root. In the lattice variables, $w_n$, these are 
\begin{align*}
y_n&= (w_{n+2}-w_n)(w_{n+2}-w_{n+1}) - a_2,\\
z_n&= \dfrac{(w_{n+2}-w_n)(y_n+a_1)}{y_n+nh},
\end{align*}
from which we extricate the second-order system
\begin{align*}
y_n + y_{n+2}  &= z_{n}^2- (a_1 + a_2),\\
z_n z_{n+2} &= -\dfrac{(y_{n+2}+a_1)(y_{n+2}+a_2)}{(y_{n+2}+(n+2)h)}.
\end{align*}
This system is equivalent to
\[
(y_n + y_{n-2} + a_1 + a_2)(y_n + y_{n+2} + a_1 + a_2) =  \dfrac{(y_n+a_1)^2(y_n+a_2)^2}{(y_n+nh)^2},
\]
which is a special version of $d$-$\mathrm{P}_{IV}$ \cite{Gramani:coalescences}. The second power of the generating shift is equivalent to the shift $(l,m) \to (l,m+1)$, hence, a simplified Lax representation for this system is
\begin{align*}
Y_n(x+2h) = A_n(x)Y_n(x),\\
Y_{n+2}(x) = B_n(x)Y_n(x),
\end{align*}
where $B_n(x) \mapsfrom M_{l,m}$. We may simplify the spectral matrix, via a gauge transformation, to be 
\begin{align*}
A_n(x) \equiv & \begin{pmatrix}  
 -\dfrac{(y_n+a_1)(y_n+a_2)}{z_n} & x^2+\delta x + \epsilon \\
 x-y_n & (y_n+nh)z_n
\end{pmatrix},
\end{align*}
where
\begin{align*}
\delta =& y_n+a_1+a_2+hn,\\ 
\epsilon =& (y_n+hn)(y_n+a_1) + (hn+y_n+a_1)a_2,
\end{align*}
and a deformation matrix, under the same transformation, becomes
\begin{align*}
B_n(x) =& \begin{pmatrix} 
 -z_{n} & x+hn+z_n^2 \\
 1 & -z_{n}
 \end{pmatrix}.
\end{align*}
Using this method, one is able to provide product formulas for the Lax representations  of $q$-$\mathrm{P}_{II}$, $q$-$\mathrm{P}_{V}$ of Hay et al. \cite{Hay} and the Lax pair for a version of $q$-$\mathrm{P}_{III}$ of Joshi et al. \cite{Ormerod:uP3}. Furthermore, this shows and explains that the Lax pairs for the reductions of \cite{Hay} factorize in a nice way. The resulting factorizations provide a simple way to compute the compatibility.

\section{Reductions of dSKdV}

We consider periodic reductions of the nonautonomous discrete Schwarzian Korteweg-de Vries equation;
\begin{align}
\label{dSlKdV}
\alpha _l &\left(\frac{1}{w_{l,m+1}-w_{l+1,m+1}}+\frac{1}{w_{l+1,m}-w_{l,m}}\right)\\
&= \beta _m \left(\frac{1}{w_{l+1,m}-w_{l+1,m+1}}+\frac{1}{w_{l,m+1}-w_{l,m}}\right)\nonumber ,
\end{align}
which possesses a Lax representation of the form \eqref{nonautLaxmats} where
\begin{subequations}\label{LaxdSlKdV}
\begin{align}
L_{l,m} &= \begin{pmatrix}
  1  & w_{l,m}-w_{l+1,m} \\
 \dfrac{\alpha_l}{ \gamma (w_{l,m}-w_{l+1,m})} & 1
 \end{pmatrix}, \\
M_{l,m} &=  \begin{pmatrix} 
 1  & w_{l,m}-w_{l,m+1} \\
 \dfrac{\beta_m}{\gamma(w_{l,m}-w_{l,m+1})} & 1  \end{pmatrix}.
\end{align}
\end{subequations}

\subsection{$q$-$\mathrm{P}(A_3^{(1)})$}

Recently, one of the authors derived a $q$-analogue of the sixth Painlev\'e equation (or $q$-$\mathrm{P}(A_3^{(1)})$) as a reduction of the non-autonomous modified Korteweg-de Vries equation \cite{Ormerod:qP6}. This work demonstrated the specified method where the transformation between the lattice variables, the $w_i$, and the Painlev\'e variables was relatively simple. We provide a similar but more complicated relation between the lattice variables for the discrete Schwarzian Korteweg-de Vries equation and the Painlev\'e variables of $q$-$\mathrm{P}(A_3^{(1)})$. 

We now provide a new reduction of the lattice equation to a version of $q$-$\mathrm{P}_{VI}$ given by
\begin{subequations}\label{qP6}
\begin{align}
y y'=&\dfrac{b_1 b_2 (z-1) \left(q^2 z-1\right)}{q^2 \left(b_1 b_2 t-\theta _1 z\right) \left(b_1 b_2 t-\theta _2 z\right)},\\
z z'=&\dfrac{\left(b_1 q^2 t y'-1\right) \left(b_2 q^2 t y'-1\right)}{q^2 \left(a_1 y'-1\right) \left(a_2 y'-1\right)}, \hspace{2cm} t' = q^2t,
\end{align}
\end{subequations}
where $\theta_1\theta_2 = a_1a_2b_1b_2$. This extends our previous result in \cite{Ormerod:qP6} by the addition of an extra parameter, which represents an integral that is used to reduce the order of the map, given by $\theta_1$ (or $\theta_2$). Obtaining a $q$-analogue of the sixth Painlev\'e equation from a discrete analogue of the Schwarzian KdV equation has some historic significance as it could be considered a discrete analogue of the reduction of the Schwarzian KdV equation to the sixth Painlev\'e equation \cite{SKdVP6I, SKdVP6II}. 

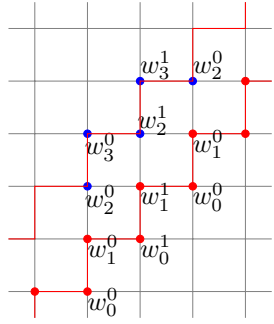
\begin{figure}[!ht]
\begin{tikzpicture}[scale=.7] 
 \draw[very thin,color=gray] (-1.5,-.5) grid (3.5,5.5);
    \draw[red] (-1,-.5) -- (-1,0) -- (0,0) -- (0,1) -- (1,1) -- (1,2) -- (2,2)-- (2,3) -- (3,3) -- (3,4) -- (3.5,4);
    \filldraw[red] (-1,0) circle(2pt);
    \filldraw[red] (0,0) circle(2pt);
    \filldraw[red] (0,1) circle(2pt);
    \filldraw[red] (1,1) circle(2pt);
    \filldraw[red] (1,2) circle(2pt);
    \filldraw[red] (2,2) circle(2pt);
    \filldraw[red] (2,3) circle(2pt);
    \filldraw[red] (3,3) circle(2pt);
    \filldraw[red] (3,4) circle(2pt);
    \begin{scope}[xshift = .3cm,yshift=-.2cm]
    \node at (0,0) {$w_0^0$};
    \node at (0,1) {$w_1^0$};
    \node at (1,1) {$w_0^1$};
    \node at (1,2) {$w_1^1$};
    \node at (2,2) {$w_0^0$};
    \node at (2,3) {$w_1^0$};    
    \end{scope}
    \node at (.25,1.75) {$w_2^0$};
    \node at (2.3,4.3) {$w_2^0$};
    \node at (1.3,4.3) {$w_3^1$};
    \node at (.25,2.75) {$w_3^0$};
    \node at (1.25,3.25) {$w_2^1$};
    \filldraw[blue] (0,2) circle(2pt);
    \filldraw[blue] (1,3) circle(2pt);
    \filldraw[blue] (2,4) circle(2pt);
    \filldraw[blue] (0,3) circle(2pt);
    \filldraw[blue] (1,4) circle(2pt);
    \begin{scope}[yshift=2cm]
    	\draw[red,thin] (-1.5,-1) -- (-1,-1) -- (-1,0) -- (0,0) -- (0,1) -- (1,1) -- (1,2) -- (2,2)-- (2,3) -- (3,3) -- (3,3.5);
    	\end{scope}
\end{tikzpicture}
\caption{The reduction and the labelling of variables. \label{qP6dSKdV}}
\end{figure}

We consider a reduction of \eqref{dSlKdV}, given by \eqref{periodicity} where $s_1 = s_2 = 2$. The labelling, given by \eqref{labelling}, is depicted in figure \ref{qP6dSKdV}. The constraint, \eqref{qseq}, becomes 
\begin{align}
\label{periodicity:qP6}\dfrac{\alpha_{l+2}}{\alpha_l} = \dfrac{\beta_{m+2}}{\beta_m} := q^2
\end{align}
which introduces the parameter $q$. 

We satisfy the periodicity constraints of \eqref{periodicity:qP6} by explicitly setting
\[
\alpha_l = \left\{\begin{array}{c p{2cm}} 
a_1 q^l & if $l$ is even,\\ 
a_2 q^l & if $l$ is odd,
  \end{array}\right. \hspace{.7cm} 
  \beta_m = \left\{\begin{array}{c p{2cm}} 
b_1 q^m & if $m$ is even,\\ 
b_2 q^m & if $m$ is odd.
  \end{array}\right.
\]
Naturally, the dependent variable, $n$, is invariant along the direction of the reduction. The correspondence between the associated linear problem here and that of Sakai \cite{SakaiE6} is made more natural by specifying an independent variable, $t$, and our spectral variable, $x$, following previous sections:
\begin{align*}
t = q^{m-l} = q^n, \hspace{.7cm} x = \dfrac{q^l}{\gamma} = \dfrac{q^k}{\gamma}.
\end{align*}
For the sake of clarity, the reduced lattice variables may be regarded as functions of $t$ or $n$ under the identification
\[
w_{n+k}^i \cong w^i(q^{k}t),
\]
so that the shift $n \to n+1$ is equivalent to the shift $t \to qt$. We know that the shift, $(l,m) \to (l+2,m+2)$, is equivalent to the shift $(x,t) \to (q^2x,t)$. Since $a=1$ and $b= 1$, the direction of our generating shift is $(l,m) \to (l,m+1)$, which is equivalent to the shift $(x,t) \to (x,q t)$. This gives us a Lax pair of the form
\begin{subequations}
\begin{align}
Y(q^2x,t) &= A(x,t)Y(x,t),\\
Y(x,qt) &= B(x,t)Y(x,t),
\end{align}
\end{subequations}
where the matrix $A(x,t)$ is given by the product
\begin{align}
A(x,t) \mapsfrom& L_{l+1,m+2}M_{l+1,m+1}L_{l,m+1}M_{l,m}, \\
\label{prodqP6}A(x,t) =& \begin{pmatrix} 1 & w_{n+1}^{1}-w_{n}^{0} \\ \dfrac{a_2 x}{w_{n+1}^{1}-w_{n}^{0}} & 1 \end{pmatrix}
\begin{pmatrix} 1 & w_{n}^{1}-w_{n+1}^{1} \\ \dfrac{t x b_2}{w_{n}^1-w_{n+1}^{1}} & 1 \end{pmatrix}\\
&
\begin{pmatrix} 1 & w_{n+1}^{0}-w_{n}^{1} \\ \dfrac{a_1 x}{w_{n+1}^{0}-w_{n}^{1}} & 1 \end{pmatrix}
\begin{pmatrix} 1 & w_{n}^{0}-w_{n+1}^{0} \\ \dfrac{t x b_1}{w_{n}^{0}-w_{n+1}^{0}} & 1\end{pmatrix}.\nonumber
\end{align}
The deformation matrix, $B(x,t)$, corresponds to the shift $m \to m+1$, hence, is given by
\begin{align}
B(x,t) \mapsfrom& M_{l,m}, \\
B(x,t)=& \begin{pmatrix} 
 1 & w_{n}^0-w_{n+1}^0 \\
 \dfrac{t x b_1}{w_{n}^{0}-w_{n+1}^{0}} & 1
\end{pmatrix}. 
\end{align}
The compatibility condition, written as
\[
A(x,qt)B(x,t) = B(q^2x,t)A(x,t),
\]
yeilds the evolution equation
\begin{subequations}\label{evolqP6}
\begin{align}
w_{n+1}^0 =& \dfrac{a_1 w_{n}^{0} \left(w_{n-1}^{1}-w_{n}^{1}\right)+t b_1 \left(w_{n}^{0}-w_{n-1}^{1}\right) w_{n}^{1}}{a_1 \left(w_{n-1}^{1}-w_{n}^{1}\right)+t b_1 \left(w_{n}^{0}-w_{n-1}^{1}\right)}, &b_1 \to \dfrac{b_2}{q},\\
w_{n+1}^1 =& \dfrac{a_2 \left(w_{n-1}^{0}-w_{n}^{0}\right) w_{n}^{1}+t b_2 w_{n}^{0}
   \left(w_{n}^{1}-w_{n-1}^{0}\right)}{a_2 \left(w_{n-1}^{0}-w_{n}^{0}\right)+t b_2 \left(w_{n}^{1}-w_{n-1}^{0}\right)}, &b_2 \to q b_1.
\end{align}
\end{subequations}
Notice that the generating shift swaps the roles of $b_1$ and $b_2$, coupled with multiplicative factors introduced to compensate for the dependence of $\beta_m$ on $q^m$. We claim that the second iterate of this mapping is $q$-$\mathrm{P}(A_3^{(1)})$. To make the full correspondence with \eqref{qP6}, let us expand out \eqref{prodqP6} to give a matrix of the form
\[
A(x,t) = A_0 + A_1x + A_2x^2,
\]
where $A_0 = I$ and the eigenvalues of $A_2$ are $\theta_1 t$ and $\theta_2 t$, where
\begin{align*}
\theta_1t &= -\dfrac{t b_1 b_2 \left(w^0_{n+1}-w^1_n\right) \left(w^0_n-w^1_{n+1}\right)}{\left(w^0_n-w^0_{n+1}\right) \left(w^1_n-w_{n+1}^1\right)},\\
\theta_2t &= -\dfrac{a_1 a_2 \left(w^0_n-w^0_{n+1}\right) \left(w^1_n-w^1_{n+1}\right)}{t \left(w^0_{n+1}-w^1_n\right) \left(w^0_n-w^1_{n+1}\right)},
\end{align*}
where $\theta_i = \theta_i(n)$ satisfies
\begin{align*}
\theta_1(n+1)\theta_1(n) = \dfrac{a_1a_2b_1b_2}{q},\\
\theta_2(n+1)\theta_2(n) = q a_1a_2b_1b_2,
\end{align*}
hence, $\theta_i(n+2) = \theta_i(n)$. That is to say that $\theta_1$ and $\theta_2$ are 2-integrals of the generating shift \cite{kintegrals}, with the additional constraint
\[
\theta_1 \theta_2  = a_1a_2b_1b_2.
\]
Secondly, from the product form, we have that
\[
\det A(x,t) = \left(a_1 x-1\right) \left(a_2 x-1\right) \left(b_1 t x-1\right) \left(b_2 t x-1\right).
\]
In accordance with the motivation given in the previous section, we choose to parameterize the spectral matrix, $A(x,t)$, in terms of the $x$-root of $(2,1)$-element of $A(x,t)$, and the diagonal entries at that root. This provides us with a spectral matrix in the curious form
\begin{equation}
A(x,t) = \begin{pmatrix}
 t x \left(x-y_n+\zeta_n\right) \theta _1+1 & t x \delta_n  \omega_n  \theta _2 \\
 \dfrac{t x (x-y_n) \theta _1}{\omega_n } & t x \left(x-y_n+\eta_n\right) \theta _2+1
\end{pmatrix},
\end{equation}
where 
\begin{align*}
\zeta_n = \dfrac{(y_na_1-1)(y_na_2-1)}{z_n}, \qquad \eta_n = (b_1ty_n-1)(b_2ty_n-1)z_n.
\end{align*}
Fixing the determinant requires that 
\[
\delta_n = \dfrac{1}{a_1}+\dfrac{1}{a_2}+\dfrac{1}{b_1 t}+\dfrac{1}{b_2 t}-2 y+\zeta_n+\eta_n.
\]
Under this identification, the $y_n$ and $z_n$ are specified by the lattice variables
\begin{subequations}
\begin{align}
\label{yndef:qP6}y_n =& - \Big[(w_{n}^1-w_{n+1}^1)((w_{n+1}^0-w_n^1)(a_2(w_{n+1}^0-w_{n}^0)+b_1 t(w_{n}^0-w_{n+1}^1))\\
&+a_1 (w_{n}^0-w_{n+1}^0)(w_{n}^0-w_{n+1}^1))+b_2 t(w_{n}^0-w_{n+1}^0)(w_{n+1}^0-w_n^1)(w_{n}^0-w_{n+1}^1)\Big]\nonumber \\
& \div \Big[b_1 t(w_{n+1}^0-w_n^1)(a_2 (w_{n+1}^0-w_{n+1}^1) (w_{n+1}^1-w_n^1)+b_2 t(w_{n+1}^0-w_n^1)\nonumber \\
& (w_{n}^0-w_{n+1}^1))-a_1 a_2 (w_{n}^0-w_{n+1}^0)(w_n^1-w_{n+1}^1)^2\Big]\nonumber \\
\label{zndef:qP6}z_n =&\Big[(w_{n}^{0}-w_{n+1}^{1})(w_{n}^{1}-w_{n+1}^{1})(a_1(w_{n+1}^{0}-w_{n}^{0})(w_{n}^{1}-w_{n+1}^{1})\\
&  -t b_2  (w_{n}^{0}-w_{n+1}^{0}) (w_{n+1}^{0}-w_{n}^{1})-b_1 t (w_{n+1}^{0}-w_{n+1}^{1})
  (w_{n+1}^{0}-w_{n}^{1}))\Big]\nonumber \\
& \div \Big[(w_{n}^{0}-w_{n+1}^{0}) (w_{n+1}^{0}-w_{n}^{1})((w_{n}^{1}-w_{n+1}^{1})(a_1 w_{n}^{0}-a_2 w_{n+1}^{0}\nonumber \\
& +(a_2-a_1) w_{n+1}^{1})+b_2 t (w_{n+1}^{0}-w_{n}^{1})(w_{n}^{0}-w_{n+1}^{1}))\Big]\nonumber .
\end{align}
\end{subequations}
We may now make the correspondence with \eqref{qP6} via the identification that $y = y_n$, $y' = y_{n+2}$, $z = z_n$ and $z' = z_{n+2}$. One may verify \eqref{qP6},  remarkably using the evolution \eqref{evolqP6} alone. We also have the gauge factor, given by
\begin{align*}
\omega_n =& -\Big[b_1 b_2 t^2 \left(w_{n+1}^{0}-w_{n}^{1}\right){}^2 \left(w_{n}^{0}-w_{n+1}^{1}\right)^2 \Big]\div \Big[b_1 t \left(w_{n+1}^{0}-w_{n}^{1}\right) \\
& \left(a_2 \left(w_{n+1}^{0}-w_{n+1}^{1}\right) \left(w_{n+1}^{1}-w_{n}^{1}\right)+b_2 t \left(w_{n+1}^{0}-w_{n}^{1}\right)
   \left(w_{n}^{0}-w_{n+1}^{1}\right)\right)\\
&-a_1 a_2 \left(w_{n}^{0}-w_{n+1}^{0}\right) \left(w_{n}^{1}-w_{n+1}^{1}\right){}^2\Big],
\end{align*}
which satisfies the equation
\begin{align*}
\dfrac{\omega_{n+2}}{\omega_n} = \dfrac{(z_n-1)(\theta _1 (z_n (a_2 y_n-1)+1)-a_2 t b_1 b_2 y_n)}{y_n (\theta_1 z_n-t b_1 b_2) (a_2^2 y_n z_n+\theta_1 t y_n-a_2 (t(b_1+b_2) y_n+z_n-1))}.
\end{align*}
We may now parameterize the deformation matrix for the double shift in terms of $y_n$, $z_n$ and $w_n$ as
\[
B(x,t) = I + \begin{pmatrix}
 \frac{x \left(\theta _1-t b_1 b_2\right)}{b_1+b_2+\left(\zeta_n-y_n\right) \theta _1} & \frac{\delta_n  \omega_n \theta _2}{b_1+b_2+\left(\zeta_n-y_n\right) \theta _1} \\
 \frac{x \left(\left(t b_1 \left(y_n-\zeta_n\right)-1\right) \theta _1-t b_1^2\right) \left(\left(t b_2 \left(y-\zeta_n\right)-1\right) \theta _1-t b_2^2\right)}{\delta_n  \omega_n  \left(\theta _1-t b_1 b_2\right) \theta _2} & \frac{x b_1 b_2
   \left(b_1+b_2+\left(\zeta_n-y_n\right) \theta _1\right) t^2}{\theta _1-t b_1 b_2}
   \end{pmatrix}.
\]
Curiously, the coefficient of $x$ is lower triangular and the constant coefficient is upper-triangular. This rather simplified Lax pair comes at the expense of the requirement that the variables $y_n$ and $z_n$ explicitly lie on the biquadratic
\begin{align*}
V(y,z,t) =& \theta _1 (z-1)^2-\theta _1 y (z-1) \left(\left(a_1+a_2\right) z-t \left(b_1+b_2\right)\right)\\
&+y^2 \left(a_1 a_2 z-\theta _1 t\right) \left(\theta _1 z-t b_1 b_2\right) = 0,
\end{align*}
which makes the computation of the compatibility condition,
\[
A(x,q^2t)B(x,t) = B(q^2x,t)A(x,t),
\]
slightly more difficult. One can verify directly that $V(y,z,t)=0$ implies that $V(y',q^2z,q^2t) = 0$ and $V(y',z', q^2 t) = 0$ under the evolution defined by \eqref{qP6}. This forms an explicit parameterization of the member of the pencil of biquadratics for each $t$. The existence of such a parameterization is not without precedent, and has appeared in the work of Yamada \cite{Yamada:LaxqEs} and Noumi et al. \cite{NoumiYamada:ellE8Lax}. We do remark that it is interesting that this explicit dependence on the bi-quadratic curve essentially came from a condition on the Lax matrices.

\subsection{$q$-$\mathrm{P}(A_2^{(1)})$ as a reduction of dSKdV}

While a reduction to $q$-$\mathrm{P}(A_3^{(1)})$ has been provided by one of the authors previously \cite{Ormerod:qP6}, we know of no reduction from an integrable lattice equation to any member of the hierarchy above $q$-$\mathrm{P}(A_3^{(1)})$. We wish to extend this further and consider reductions of a Painlev\'e equation that is higher up in the classification scheme; namely the $q$-Painlev\'e equation with $E_6^{(1)}$ symmetry, which is associated with a surface with $A_2^{(1)}$ symmetry, (or $q$-$\mathrm{P}(A_2^{(1)})$), given by \eqref{qPE6}. We will find that this equation appears as a reduction of \eqref{dSlKdV}, or $Q1_m^{\delta=0}$, or the non-autonomous discrete Schwarzian Korteweg-de Vries equation \cite{Nijhoff:dSKdV}.

\begin{figure}[!ht]
\begin{tikzpicture}[scale=.7] 
 \draw[very thin,color=gray] (-1.5,-.5) grid (5.5,5.5);
    \draw[red] (-1.5,0) -- (-1,0) -- (0,0) -- (0,1) -- (1,1) -- (2,1) -- (2,2)-- (3,2) -- (4,2) -- (4,3) -- (5.5,3);
    \begin{scope}[yshift=2cm]
    \draw[red] (-1.5,0) -- (-1,0) -- (0,0) -- (0,1) -- (1,1) -- (2,1) -- (2,2)-- (3,2) -- (4,2) -- (4,3) -- (5.5,3);
    \end{scope}
    \filldraw[red] (-1,0) circle(2pt);
    \filldraw[red] (0,0) circle(2pt);
    \filldraw[red] (0,1) circle(2pt);
    \filldraw[red] (1,1) circle(2pt);
    \filldraw[red] (2,1) circle(2pt);
    \filldraw[red] (2,2) circle(2pt);
    \filldraw[red] (3,2) circle(2pt);
    \filldraw[red] (4,2) circle(2pt);
    \filldraw[red] (4,3) circle(2pt);
    \filldraw[red] (5,3) circle(2pt);
    \begin{scope}[xshift = .3cm,yshift=-.3cm]
	\node at (-1,0) {$w_1^0$};
    \node at (0,0) {$w_0^1$};
    \node at (0,1) {$w_2^0$};
    \node at (1,1) {$w_1^1$};
    \node at (2,1) {$w_0^0$};
    \node at (2,2) {$w_2^1$};
    \node at (3,2) {$w_1^0$};    
    \node at (4,2) {$w_0^1$};    
    \node at (4,3) {$w_2^0$}; 
    \node at (5,3) {$w_1^1$}; 
    \end{scope}
    \filldraw[blue] (-1,1) circle(2pt);
    \filldraw[blue] (-1,2) circle(2pt);
    \filldraw[blue] (1,2) circle(2pt);
    \filldraw[blue] (3,3) circle(2pt);
    \filldraw[blue] (0,2) circle(2pt);
    \filldraw[blue] (0,3) circle(2pt);
    \filldraw[blue] (1,3) circle(2pt);
    \filldraw[blue] (2,3) circle(2pt);
    \filldraw[blue] (2,4) circle(2pt);
    \filldraw[blue] (3,4) circle(2pt);
    \filldraw[blue] (4,4) circle(2pt);
    \filldraw[blue] (4,5) circle(2pt);
    \filldraw[blue] (5,5) circle(2pt);
    \filldraw[blue] (5,4) circle(2pt);
    \begin{scope}[xshift = .3cm,yshift=1.7cm]
	\node at (-1,0) {$w_5^0$};
    \node at (0,0) {$w_4^1$};
    \node at (-1,-1) {$w_3^1$};
    \node at (5,2) {$w_3^0$};
    \node at (0,1) {$w_6^0$};
    \node at (3,1) {$w_3^1$};
    \node at (1,0) {$w_3^0$};
    \node at (1,1) {$w_5^1$};
    \node at (2,1) {$w_4^0$};
    \node at (2,2) {$w_6^1$};
    \node at (3,2) {$w_5^0$};    
    \node at (4,2) {$w_4^1$};    
    \node at (4,3) {$w_6^0$}; 
    \node at (5,3) {$w_5^1$}; 
    \end{scope}
\end{tikzpicture}
\caption{The reduction and the labelling of variables.\label{figred}}
\end{figure}
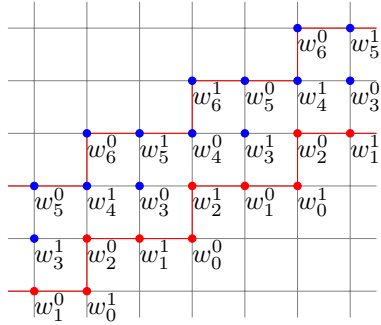
We impose \eqref{periodicity} on \eqref{dSlKdV} with $s_1 =4$ and $s_2 = 2$, hence, our constaint, \eqref{qseq}, becomes
\[
\dfrac{\alpha_{l+4}}{\alpha_l} = \dfrac{\beta_{m+2}}{\beta_m} := q^4,
\]
which we solve in a similar manner as before;
\begin{equation}\label{parameterquasiperiodicity}
\alpha_l = \left\{ \begin{array}{l p{2.5cm}}
a_1q^l & if $l = 0$ mod $4$\\
a_2q^l & if $l = 1$ mod $4$\\
a_3q^l & if $l = 2$ mod $4$\\
a_4q^l & if $l = 3$ mod $4$
\end{array} \right., \hspace{.5cm} \beta_m = \left\{ \begin{array}{l p{2.5cm}}
b_1q^{2m} & if $m = 0$ mod $2$\\
b_2q^{2m} & if $m = 1$ mod $2$
\end{array} \right. .
\end{equation}
We use the same Lax pair for \eqref{dSlKdV} as the previous section, namely \eqref{LaxForm} where $L_{l,m}$ and $M_{l,m}$ are specified by \eqref{LaxdSlKdV}. The $t$-direction is also chosen to be constant in the direction of the reduction. We choose our spectral parameter, $x$, and independent variable, $t$, in a manner in which correspondence with $q$-$\mathrm{P}(A_2^{(1)})$ comes more naturally, that is
\begin{align*}
t = q^{2m-l} = q^n, \hspace{.7cm} x = \dfrac{q^k}{\gamma} = \dfrac{q^l}{\gamma}.
\end{align*}
We note that the generating shift is no longer just a shift in $m$ alone, but a simultaneous shift in $l$ and $m$ , i.e., $(l,m) \to (l+1,m+1)$, hence we present a Lax pair that represents the shift $(l,m) \to (l+4,m+2)$, which is now equivalent to $(x,t) \to (q^4x,t)$, and the shift $(l,m) \to (l+1,m+1)$, which is equivalent to the shift $(x,t) \to (qx,qt)$. Hence, our Lax pair is a linear system satisfying
\begin{subequations}\label{Lax:E6}
\begin{align}
Y(q^4x,t) &= A(x,t)Y(x,t),\\
Y(qx,qt) &= B(x,t)Y(x,t),
\end{align}
\end{subequations}
where we compose $A$ and $B$ in terms of $L$ and $M$ in the following way
\begin{subequations}\label{productE6}
\begin{align}
A(x,t) \mapsfrom& L_{l+3,m+2}L_{l+2,m+2}M_{l+2,m+1}L_{l+1,m+1}L_{l,m+1}M_{l,m},\\
\label{prodA:E6}A(x,t)=& 
\begin{pmatrix} 1 & w_{n+1}^0-w_{n}^1 \\
 \dfrac{x a_4}{w_{n+1}^{0}-w_{n}^{1}} & 1 \end{pmatrix} 
\begin{pmatrix} 1 & w_{n+2}^{1}-w_{n+1}^{0} \\
 \dfrac{x a_3}{w_{n+2}^{1}-w_{n+1}^{0}} & 1\end{pmatrix} \\
&\begin{pmatrix}  1 & w_{n}^{0}-w_{n+2}^{1} \\
 \dfrac{t x b_2}{w_{n}^{0}-w_{n+2}^{1}} & 1\end{pmatrix} 
\begin{pmatrix} 1 & w_{n+1}^{1}-w_{n}^{0} \\
 \dfrac{x a_2}{w_{n+1}^{1}-w_{n}^{0}} & 1 \end{pmatrix} \nonumber \\
&\begin{pmatrix} 1 & w_{n+2}^{0}-w_{n+1}^{1} \\
 \dfrac{x a_1}{w_{n+2}^{0}-w_{n+1}^{1}} & 1 \end{pmatrix} 
\begin{pmatrix}  1 & w_{n}^{1}-w_{n+2}^{0} \\
 \dfrac{t x b_1}{w_{n}^{1}-w_{n+2}^{0}} & 1 \end{pmatrix}. \nonumber 
\end{align}
The deformation matrix may be written as
\begin{align}
B(x,t) &\mapsfrom L_{l,m+1}M_{l,m},\\
B(x,t)&= \begin{pmatrix} 1 & w_{n+2}^{0}-w_{n+1}^{1} \\
 \dfrac{x a_1}{w_{n+2}^{0}-w_{n+1}^{1}} & 1 \end{pmatrix} 
\begin{pmatrix}  1 & w_{n}^{1}-w_{n+2}^{0} \\
 \dfrac{t x b_1}{w_{n}^{1}-w_{n+2}^{0}} & 1 \end{pmatrix},
\end{align}
\end{subequations}
where we have written these as functions of $x$ and $t$. The compatibility of the system defined by \eqref{Lax:E6} is given by
\begin{equation}\label{CompE6}
A\left(q x, q t \right)B(x,t) = B(q^4x,t)A(x,t).
\end{equation}
One may use \eqref{CompE6} to derive the required evolution equations; the equations defining the evolution may be written as
\begin{subequations}\label{EvolE6}
\begin{align}
w_{n+1}^0 = \frac{a_4 \left(w_{n-2}^0-w_n^1\right) w_{n-1}^1+b_2 t w_n^1 \left(w_{n-1}^1-w_{n-2}^0\right)}{a_4 \left(w_{n-2}^0-w_n^1\right)+b_2 t
   \left(w_{n-1}^1-w_{n-2}^0\right)},\\
w_{n+1}^1 = \frac{a_2 w_{n-1}^{0} \left(w_{n-2}^{1}-w_{n}^{0}\right)+b_1 t \left(w_{n-1}^{0}-w_{n-2}^{1}\right) w_{n}^{0}}{w_{n-2}^{1} \left(a_2-b_1 t\right)-a_2 w_{n}^{0}+b_1 t w_{n-1}^{0}},\\
a_1 \to \dfrac{a_2}{q}, \hspace{.7cm} a_2 \to \dfrac{a_3}{q}, \hspace{.7cm} a_3 \to \dfrac{a_4}{q}, \hspace{.7cm} a_4 \to a_1 q^3,\\
b_1 \to \dfrac{b_2}{q^2}, \hspace{1cm} b_2 \to q^2 b_1. \hspace{2cm}
\end{align}
\end{subequations}
The difficult step is to extract a second order system from this seemingly sixth order system via some special parameterization. To do this, we observe the properties of \eqref{Lax:E6} as an associated linear problem. Firstly, expanding out \eqref{prodA:E6} in the spectral parameter, we find 
\[
A(x,t) = A_0 + A_1x + A_2x^2 + A_3 x^3,
\]
where $A_0$ is the identity matrix and $A_3$ is a lower triangular matrix with diagonal entries $\theta_1 t$ and $\theta_2 t$, where
\begin{align*}
\theta_1  = \frac{a_2 a_3 b_1 \left(w_{n}^{1}-w_{n+1}^{0}\right) \left(w_{n+2}^{0}-w_{n+1}^{1}\right) \left(w_{n}^{0}-w_{n+2}^{1}\right)}{\left(w_{n}^{0}-w_{n+1}^{1}\right) \left(w_{n+2}^{0}-w_{n}^{1}\right) \left(w_{n+1}^{0}-w_{n+2}^{1}\right)},\\
\theta_2  = \frac{a_1 a_4 b_2 \left(w_{n}^{0}-w_{n+1}^{1}\right) \left(w_{n}^{1}-w_{n+2}^{0}\right) \left(w_{n+1}^{0}-w_{n+2}^{1}\right)}{\left(w_{n+1}^{0}-w_{n}^{1}\right) \left(w_{n+2}^{0}-w_{n+1}^{1}\right) \left(w_{n}^{0}-w_{n+2}^{1}\right)},
\end{align*}
where, $\theta_1$ and $\theta_2$ are invariants in accordance with the evolution equations, given by \eqref{EvolE6}. The simplicity of the individual factors of $A(x,t)$ from \eqref{prodA:E6} give us that the determinant is
\begin{equation}\label{detAE6}
\det A(x,t) = \left(x a_1-1\right) \left(x a_2-1\right) \left(x a_3-1\right) \left(x a_4-1\right) \left(tx b_1-1\right) \left(tx b_2-1\right).
\end{equation}
These properties should remind us of the properties of a special case of the spectral matrix in the Lax pair of Sakai \cite{SakaiE6}. We may simply use a transformation of the form $Y(x,t) \to S Y(x,t)$, where $S$ is a lower triangular matrix that is constant in $x$ such that the transformation $A \to SAS^{-1}$ diagonalises $A_3$. While this matrix, $S$, isn't particularly nice to write down, the resulting matrix is in the general form
\[
A(x,t) = x\begin{pmatrix}
\theta _1 t  \left((x-y_n) (x-\epsilon_n )+\zeta _n\right) & \theta _2 t \omega_n  (x-y_n)  \\
\dfrac{\theta _1t  (x \gamma_n +\delta_n ) }{\omega_n} & \theta_2 t  \left((x-y_n) (x-\chi_n )+\eta_n\right) 
 \end{pmatrix} + I.
\]
The terms $\epsilon_n$, $\chi_n$, $\gamma_n$ and $\delta_n$ may be determined from \eqref{detAE6} in terms of $y_n$ and $\zeta_n$, $\eta_n$, the $a_i$'s and the $b_i$'s. The relation between this form of $A(x,t)$ and the known Lax pairs of Sakai \cite{SakaiE6} and Yamada \cite{Yamada:LaxqEs} has recently been found by one of the authors \cite{Ormerod:qE6}. Following the motivation from the previous section, this parameterization defines $y_n$ to be the root of the $(1,2)$-entry, leaving a choice of $\zeta_n$ and $\eta_n$ such that
\[
(\theta_1 t \zeta _n y_n+1)(\theta_2 t \eta _n y_n+1)= \det A(y_n,t),
\]
where the right hand side is defined by \eqref{detAE6}. With this in mind the defining equations for $\zeta_n$ and $\eta_n$ are
\begin{align*}
\theta_1 t \zeta _n y_n+1&=(1-y_n z_n) \left(b_1 t y_n-1\right) \left(b_2 t y_n-1\right),\\
\theta_2 t \eta_n y_n+1&=\dfrac{\left(a_1 y_n-1\right) \left(a_2 y_n-1\right) \left(a_3 y_n-1\right) \left(a_4 y_n-1\right)}{(1-y_n z_n)}.
\end{align*}
This specifies $y_n$ and $z_n$ in terms of the lattice variables:
\begin{align}
y_n =& \Big[(w_{n+2}^0-w_{n+1}^1) ((w_{n}^0-w_{n+1}^1) (w_{n}^1-w_{n+2}^1) (b_2 t (w_{n}^0-w_{n}^1) (w_{n+1}^0-w_{n+2}^1)\\ 
& -a_3 (w_{n+1}^0-w_{n}^1) (w_{n}^0-w_{n+2}^1))+a_2
   (w_{n}^0-w_{n}^1) (w_{n}^1-w_{n+1}^1) (w_{n}^0-w_{n+2}^1)\nonumber\\
   & (w_{n+2}^1-w_{n+1}^0))+a_1 (w_{n}^0-w_{n+1}^1) (w_{n}^1-w_{n+1}^1) (w_{n+2}^0-w_{n}^1)
   (w_{n}^0-w_{n+2}^1)\nonumber\\
 & (w_{n+1}^0-w_{n+2}^1)\Big]\div\Big[a_1 (w_{n}^0-w_{n+1}^1) (w_{n+2}^0-w_{n}^1) (a_3 (w_{n+1}^0-w_{n}^1)\nonumber\\
 &  (w_{n}^0-w_{n+2}^1)(w_{n+2}^1-w_{n+1}^1)+b_2 t
   (w_{n}^0-w_{n+1}^1) (w_{n}^1-w_{n+2}^1) \nonumber\\
 &(w_{n+1}^0-w_{n+2}^1))-a_2 a_3 (w_{n+1}^0-w_{n}^1) (w_{n}^1-w_{n+1}^1) (w_{n+2}^0-w_{n+1}^1) \nonumber\\
 & (w_{n}^0-w_{n+2}^1)^2\Big],\nonumber 
\end{align}
\begin{align} 
z_n=& [a_1 (w_{n}^0-w_{n+1}^1) ((w_{n+2}^0-w_{n+2}^1) (a_3 (w_{n+1}^0-w_{n}^1) (w_{n}^0-w_{n+2}^1) (w_{n+2}^1-w_{n+1}^1)\\
&+b_2 t (w_{n}^0-w_{n+1}^1) (w_{n}^1-w_{n+2}^1)
   (w_{n+1}^0-w_{n+2}^1))-a_2 (w_{n+2}^0-w_{n+1}^1)\nonumber\\
& (w_{n}^0-w_{n+2}^1) (w_{n+2}^1-w_{n}^1) (w_{n+2}^1-w_{n+1}^0))+a_2 a_3 (w_{n+1}^0-w_{n}^1) (w_{n+2}^0-w_{n+1}^1)\nonumber\\
&(w_{n+1}^1-w_{n+2}^1) (w_{n}^0-w_{n+2}^1){}^2]\div[(w_{n}^0-w_{n+2}^1) (w_{n+2}^1-w_{n+1}^0) ((w_{n}^0-w_{n+1}^1) \nonumber\\
&(a_1 (w_{n+2}^0-w_{n}^1) (w_{n+1}^1-w_{n+2}^1)-b_2 t
   (w_{n+2}^0-w_{n+1}^1) (w_{n}^1-w_{n+2}^1))\nonumber\\
 &+a_2 (w_{n}^1-w_{n+1}^1) (w_{n+2}^0-w_{n+1}^1) (w_{n}^0-w_{n+2}^1))]\nonumber,
\end{align}
which satisfy \eqref{ybar} and \eqref{zbar} under the identification of $y = y_n$, $z = z_n$, $y' = y_{n+4}$ and $z' = z_{n+4}$. While these expressions may fail to be succinct, they do succeed in being very explicit. Furthermore, using this identification, it is possible to directly verify \eqref{ybar} and \eqref{zbar} from \eqref{EvolE6} alone.

\section{Conclusion}

Given a nonautonomous partial difference equation which admits an additive or multiplicative form, \eqref{q:add} or \eqref{q:mult}, we have outlined a direct method for finding a Lax representation for any periodic (travelling wave) reduction of the form \eqref{periodicity}. The method outlines how the Lax matrices may be expressed in terms of products of the Lax matrices of the partial difference equation from which it was derived. 

We have shown that the method applies to deautonomized versions of the equations in the ABS list, with the exception of Q3 and Q4. We have concentrated on two cases, the discrete Korteweg-de Vries equation and the discrete Schwarzian Korteweg-de Vries equation. We have also found the relation between the derived reductions and discrete Painlev\'e equations d-$\mathrm{P}_I$ and d-$\mathrm{P}_{IV}$, respectively reductions from the discrete Schwarzian Korteweg-de Vries equation and $q$-$\mathrm{P}_{VI}$ and $q$-$\mathrm{P}(A_2^{(1)})$. The latter is, to our best knowledge, the highest full parameter member of the Sakai classification derived as a reduction so far. 

\section*{Acknowledgments}

This research is supported by Australian Research Council Discovery Grant \#DP110100077.

\bibliography{C:/Mathematics/TeX/refs}{}
\bibliographystyle{plain}

\end{document}